\documentclass[a4paper,fleqn,usenatbib]{mnras}

\usepackage{newtxtext,newtxmath}

\usepackage[T1]{fontenc}
\usepackage{ae,aecompl}

\usepackage{graphicx}	
\usepackage{amsmath}	

\newcommand{\vsini}{$v_{\rm e} \sin i$}
\newcommand{\kms}{km\,s$^{-1}$}
\newcommand{\ms}{m\,s$^{-1}$}
\newcommand{\rsun}{$R_\odot$}
\newcommand{\msun}{$M_\odot$}

\newcommand{\teff}{$T_{\rm eff}$}
\newcommand{\logg}{$\log{g}$}
\newcommand{\cas}{V772~Cas}

\newcommand{\figps}[3]{\resizebox{#1}{!}{\rotatebox{#2}{\includegraphics{#3}}}}

\usepackage{color}

\title[Eclipsing binary V772~Cas]{V772~Cas: an ellipsoidal HgMn star in an eclipsing binary}

\author[O. Kochukhov et al.]
{O.\ Kochukhov$^1$\thanks{E-mail: oleg.kochukhov@physics.uu.se},
C.\ Johnston$^2$,
J.\ Labadie-Bartz$^{3}$,
S.\ Shetye$^{4}$,
T.\ A.\ Ryabchikova$^5$,
\newauthor{
A.\ Tkachenko$^2$,
M.\ E.\ Shultz$^6$
}\\
$^1$Department of Physics and Astronomy, Uppsala University, Box 516, Uppsala 75120, Sweden \\
$^2$Instituut voor Sterrenkunde, KU Leuven, Celestijnenlaan 200D, 3001, Leuven, Belgium\\
$^3$Instituto de Astronomia, Geof\'isica e Ciencias Atmosf\'ericas, Universidade de S\`ao Paulo, Rua do Mat\~ao 1226,\\
Cidade Universit\~aria, S\'ao Paulo, SP 05508-900, Brazil\\
$^4$Institute of Astronomy and Astrophysics (IAA), Université Libre de Bruxelles (ULB), CP 226, Boulevard du Triomphe,1050 Bruxelles, Belgium\\
$^5$Institute of Astronomy, Russian Academy of Sciences, Pyatnitskaya 48, 119017 Moscow, Russia\\
$^6$Department of Physics and Astronomy, University of Delaware, 217 Sharp Lab, Newark, Delaware, 19716, USA\\
\\}

\date{Accepted XXX. Received YYY; in original form ZZZ}

\pubyear{2020}

\begin{document}
\label{firstpage}
\pagerange{\pageref{firstpage}--\pageref{lastpage}}
\maketitle

\begin{abstract}
The late B-type star V772~Cas (HD\,10260) was previously suspected to be a rare example of a magnetic chemically peculiar star in an eclipsing binary system. Photometric observations of this star obtained by the TESS satellite show clear eclipses with a period of 5.0137~d accompanied by a significant out-of-eclipse variation with the same period. High-resolution spectroscopy reveals V772~Cas to be an SB1 system, with the primary component rotating about a factor two slower than the orbital period and showing chemical peculiarities typical of non-magnetic HgMn chemically peculiar stars. This is only the third eclipsing HgMn star known and, owing to its brightness, is one of the very few eclipsing binaries with chemically peculiar components accessible to detailed follow-up studies. Taking advantage of the photometric and spectroscopic observations available for V772~Cas, we performed modelling of this system with the {\sc PHOEBE} code. This analysis provided fundamental parameters of the components and demonstrated that the out-of-eclipse brightness variation is explained by the ellipsoidal shape of the evolved, asynchronously rotating primary. This is the first HgMn star for which such variability has been definitively identified.
\end{abstract}




\section{Introduction}
\label{sec:intro}

About 10 per cent of A and B main sequence stars possess stable, globally organised magnetic fields with a strength of at least 300~G \citep{auriere:2007,sikora:2019a}. These stars typically exhibit anomalous absorption spectra, shaped by surface overabundances of Si, Fe-peak, and rare-earth elements, and are known as magnetic chemically peculiar or ApBp stars. The process of radiatively driven atomic diffusion \citep{michaud:2015} responsible for these abundance anomalies also produces a high-contrast, long-lived non-uniform horizontal distribution of chemical elements. Local variation of metal abundances associated with these chemical spots modifies emergent stellar radiation and leads to a characteristic photometric rotational modulation known as $\alpha^2$~CVn (ACV) type of stellar variability \citep{samus:2017}.

The properties of magnetic fields of ApBp stars do not depend on stellar mass or rotation rate, making them strikingly different from characteristics of the dynamo-generated fields observed in late-type stars \citep[e.g.][]{vidotto:2014}. It is believed that the strong fields found in early-type stars are dynamically stable, `fossil' remnants of the magnetic flux generated or acquired by these stars at some earlier evolutionary phase \citep{braithwaite:2004,neiner:2015}. The origin of fossil fields is not understood. This magnetic flux might be inherited from molecular clouds at stellar birth \citep{mestel:1999}, produced by the convective dynamo during pre-main sequence evolution \citep{moss:2004} or created by a short-lived dynamo operating during stellar merger events \citep{schneider:2019}.

The binary characteristics of early-type magnetic stars may provide crucial clues, allowing one to test alternative fossil field hypotheses. The non-magnetic chemically peculiar stars of Am (A-type stars with enhanced lines of Fe-peak elements) and HgMn (late-B stars identified by strong lines of Hg and/or Mn) types are frequently found in close binaries (\citealt*{gerbaldi:1985}; \citealt{ryabchikova:1998a}; \citealt{carquillat:2007}), including eclipsing systems \citep{nordstrom:1994,strassmeier:2017,takeda:2019}. In contrast, only about ten close ($P_{\rm orb}<20$~d) spectroscopic binaries containing at least one magnetic ApBp star are known \citep{landstreet:2017}. The overall incidence rate of magnetic upper main sequence stars in close binaries is less than 2 per cent \citep{alecian:2015a}, although this fraction is significantly higher if one includes wide long-period systems \citep{mathys:2017}. This low incidence of magnetic ApBp stars in close binaries is frequently considered as an argument in favour of the stellar merger origin of fossil fields \citep{de-mink:2014,schneider:2016}. In this context, confirmation of magnetic ApBp stars in short-period binary systems gives support to alternative theories or, at least, demonstrates that early-type stars may acquire magnetic fields through different channels. In addition, detached close binary stars, particularly those showing eclipses, are valuable astrophysical laboratories that provide model independent stellar parameters and allow one to study pairs of co-evolving stars formed in the same environment. Until recently, no early-type magnetic stars in eclipsing binaries were known. The first such system, HD\,66051, was identified by \citet{kochukhov:2018b}. The second system, HD\,62658 containing twin components of which only one is magnetic, was found by \citet{shultz:2019}. Several other eclipsing binaries containing candidate ApBp stars were proposed (\citealt{hensberge:2007}; \citealt*{gonzalez:2010a}; \citealt{skarka:2019}), but the magnetic nature of these stars has not been verified by direct detections of their fields using the Zeeman effect. In this paper we put a spotlight on another candidate eclipsing magnetic Bp star, which received little attention prior to our work despite being significantly brighter than the confirmed magnetic eclipsing systems HD\,62658 and HD\,66051.

\cas\ (HR\,481, HD\,10260, HIP\,7939) is a bright ($V=6.7$) but little studied chemically peculiar late-B star. The exact type of its spectral peculiarity is uncertain. \citet{cowley:1972} classified this star as B8IIIpSi whereas \citet{dworetsky:1976} considered it an HgMn star. Both studies used low-dispersion classification spectra. The former BpSi classification appears to be more common in recent literature \citep[e.g.][]{gandet:2008,renson:2009,skarka:2019}. The variable star designation comes from \citet{kazarovets:1999}, who suggested this object to be an $\alpha^2$~CVn-type variable based on the Hipparcos epoch photometry \citep{perryman:1997}. \citet{otero:2007} discovered eclipses in the Hipparcos light curve. This analysis was improved by \citet{gandet:2008}. He confirmed the presence of primary eclipses, derived an orbital period of 5.0138~d and demonstrated that archival photographic radial velocity measurements \citep{hube:1970} show coherent variation with the same period. These results, along with the B8IIIpSi spectral classification and an evidence of the out-of-eclipse variability, led \citet{gandet:2008} to suggest \cas\ as an $\alpha^2$~CVn variable in a short-period eclipsing binary -- an exceptionally rare and interesting object akin to the recently discovered magnetic eclipsing binaries HD\,66051 and HD\,62658. Apart from the study by \citet*{huang:2010}, who determined \teff\,=\,$13188\pm250$~K and \logg\,=\,$3.43\pm0.05$ from the low-resolution hydrogen Balmer line spectra, no model atmosphere and/or abundance analysis was carried out for \cas. In fact, to the best of our knowledge, this star was never studied with high-resolution spectra.

In this paper we present a detailed photometric and spectroscopic investigation of \cas\ that provides a new insight into the nature of this star. In Sect.~\ref{sec:obs} we describe the new observational data used in our study. Orbital modelling and derivation of the binary component parameters are presented in Sect.~\ref{sec:binary}. Model atmosphere parameters and chemical abundances of the primary are determined in Sect.~\ref{sec:primary}. The paper concludes with the discussion in Sect.~\ref{sec:disc}.

\section{Observations}
\label{sec:obs}

\subsection{Space photometry}

The NASA Transiting Exoplanet Survey Satellite \citep[TESS;][]{ricker:2015} began its nominal 2 year mission in 2018 to discover Earth-sized transiting exoplanets. Using four cameras that cover a combined field of view of 24$^{\circ}$ $\times$ 96$^{\circ}$ and a red filter that records light in the range 6000 to 10500\AA, both ecliptic hemispheres are surveyed for one year each, in 13 sectors that extend from the ecliptic plane to the pole. Each sector is observed for approximately 27.5 days, and a given field on the sky can be observed in multiple sectors if it falls on overlap regions. Nominal TESS targets are bright, with $I_c$ magnitudes between 4 and 13, with a noise floor of approximately 60 ppm hr$^{-1}$.

Full Frame Images (FFIs) from TESS are available at a 30-minute cadence for the entire field of view, allowing light curves to be extracted for all objects that fall on the detector. Certain high priority targets were pre-selected by the TESS mission to be observed with 2-minute cadence, some of which were chosen from guest investigator programs.

V772 Cas was observed in cycle 2 of the TESS mission in sector 18 (2019-Nov-02 to 2019-Nov-27) in 30-minute cadence mode (and was not pre-selected for 2-minute cadence observations; TESS Input Catalog ID: 444833007). A target pixel file of a 50 $\times$ 50 grid of pixels centered on V772 Cas was downloaded with TESScut\footnote{\url{https://mast.stsci.edu/tesscut/}} \citep{Brasseur2019}, with further processing aided by the \textsc{lightkurve} package \citep{Lightkurve2018}.
A light curve for \cas\ was extracted with aperture photometry using an aperture of 36 pixels centred on the target. This aperture was chosen to minimise contaminating flux from neighbouring stars, which would otherwise bias the binary system modelling and possibly introduce additional signals from other sources, while still achieving a high signal-to-noise ratio. A principal component analysis routine was then applied to the extracted light curve to detrend against signals common to neighbouring pixels (outside of a 15 $\times$ 15 pixel exclusion zone). After removing outliers and data points more prone to systematic effects (mostly scattered light near TESS orbital perigee), the light curve includes 958 observations spanning 22.1 days. 

The Hipparcos light curve analysed by \citet{otero:2007} and \citet{gandet:2008} covered 233 orbital cycles with about a dozen photometric data points tracing the primary eclipse. Although the TESS data corresponds to a time span of just 4.4 orbits, it samples the eclipses much more densely owing to its higher measurement cadence. Moreover, individual TESS measurements are about a factor of 200 more precise than the Hipparcos photometry.

\subsection{High-resolution spectroscopy}

We obtained high-resolution spectra of \cas\ using the High-Efficiency and high-Resolution Mercator \'Echelle Spectrograph (HERMES) mounted on the 1.2 m Mercator telescope at the Observatorio del Roque de los Muchachos, La Palma, Canary Islands, Spain. This instrument provides coverage of the 3700--9100~\AA\ wavelength region at a resolving power of 85\,000 \citep{raskin:2011}. The star was observed on five consecutive nights, from January 27 to January 31, 2020, with two spectra obtained on each night. Exposure times of 600--800~s were used, yielding a signal-to-noise ratio ($S/N$) of 190--280 per $\approx$\,0.03~\AA\ pixel of the extracted spectrum in the wavelength interval from 5000 to 5500~\AA. The HERMES pipeline reduction software was employed to perform the basic \'echelle data reduction steps, including the bias and flat field corrections, extraction of one-dimensional spectra and wavelength calibration. The resulting merged, un-normalised spectra were then normalised to the continuum with the methodology described by \citet{rosen:2018}.

The log of ten HERMES observations of \cas\ is given in Table~\ref{tab:obs}. The first four columns of this table list the UT date of observation, the corresponding heliocentric Julian date, the orbital phase calculated according to the ephemeris derived in Sect.~\ref{sec:binary}, and the $S/N$ ratio. The last column provides radial velocities determined in the next section.

\subsection{Spectroscopic classification and radial velocity measurements}

Initial qualitative analysis of the high-resolution spectra of \cas\ showed the presence of a single set of spectral lines with variable radial velocity and line strengths typical of HgMn late-B chemically peculiar stars. Specifically, the HgMn classification of this star is unambiguously demonstrated by the presence of \ion{Hg}{ii} 3984~\AA, \ion{Ga}{ii} 6334~\AA\ and numerous strong \ion{Mn}{ii} and \ion{P}{ii} lines. On the other hand, ionised Si lines are not anomalously strong and rare-earth absorption features are not prominent in the spectrum of \cas, which argues against identification of this object as a Si-peculiar Bp star.

\begin{figure}
\centering
\figps{0.85\hsize}{0}{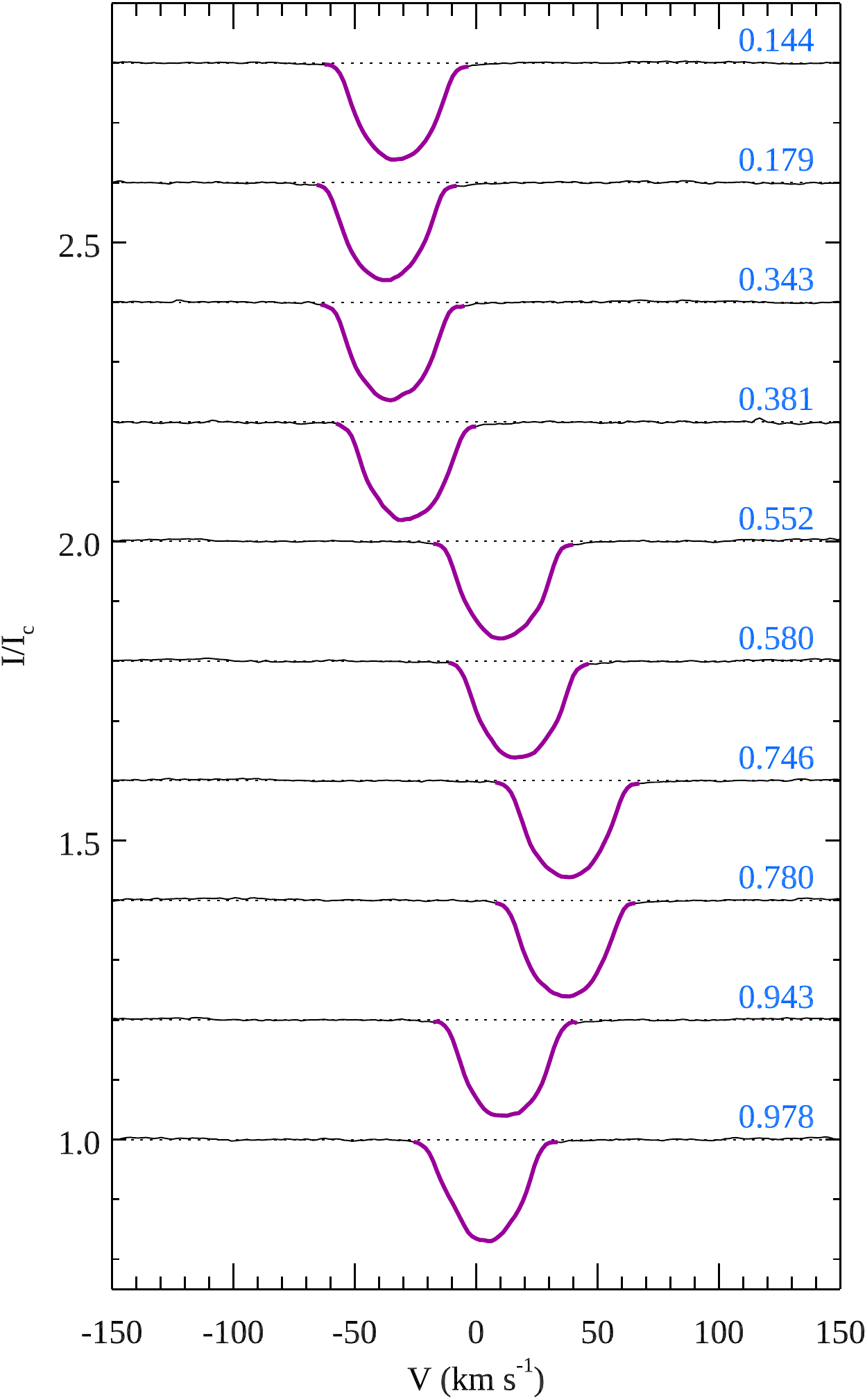}
\caption{Least-squares deconvolved profiles of V772 Cas. Profiles for different orbital phases are shifted vertically with a step of 0.2. The dotted lines show continuum level for each spectrum. The solid lines show LSD profile for the entire velocity span with the thicker segments indicating the velocity internal employed for radial velocity measurements. The orbital phase corresponding to each observation is given on the right.}
\label{fig:lsd}
\end{figure}

Based on this initial assessment, we extracted a line list from the VALD database (\citealt{ryabchikova:2015}; \citealt*{pakhomov:2019}) with the chemical abundances typical of HgMn stars \citep*{ghazaryan:2018} and model atmosphere parameters \teff\,=\,13000~K, \logg\,=\,3.5, close to the values reported by \citet{huang:2010}\footnote{Precise choice of atmospheric parameters and abundances is not important for the multi-line method used in this paper.}. This line list was employed for calculation of least-squares deconvolved (LSD) profiles \citep*{kochukhov:2010a} with the goal to study the radial velocity variation of the primary and search for spectral signatures of the secondary. The LSD profiles were constructed by combining 1094 metal lines deeper than 5 per cent of the continuum. These average spectra are illustrated in Fig.~\ref{fig:lsd}, where the data are phased with the orbital ephemeris $HJD=2458803.4016+5.0138\times E$ determined below (Sect.~\ref{sec:binary}). 

The radial velocity of the primary was measured from the LSD profiles with the help of the centre-of-gravity method. These measurements made use of the $\pm30$~\kms\ velocity range around the line centre.  The resulting radial velocity changes from $-37$ to 38~\kms. This variation is coherent on the 5~d time-scale covered by the spectroscopic observations and is consistent with the orbital period seen in photometry. Individual radial velocities are reported in the last column of Table~\ref{tab:obs}. The formal uncertainty of these measurements is 50--60~\ms. 

At the same time, no evidence of a spectral contribution of the secondary was found in the LSD profiles. This is not surprising considering the luminosity ratio of $>$\,100 estimated from the eclipse depths in the TESS light curve. We tested alternative solar-composition LSD line masks for \teff\ of 7000, 9000, and 11000~K but still could not detect any features in the LSD profile time series that could be attributed to the secondary component.

\begin{table}
\caption{Log of spectroscopic observations of \cas. \label{tab:obs}}
 \begin{tabular}{llllr}
\hline\hline
UT date & HJD & Orbital & $S/N$ & $V_{\rm r}~~~~$ \\
        &     & phase & (pixel$^{-1}$)           & (\kms) \\
\hline
2020-01-27 & 2458876.3638 & 0.552 & 256 &  $ 11.17\pm0.05$ \\
2020-01-28 & 2458876.5045 & 0.580 & 237 &  $ 17.54\pm0.06$ \\  
2020-01-28 & 2458877.3365 & 0.746 & 223 &  $ 37.70\pm0.06$ \\  
2020-01-29 & 2458877.5078 & 0.780 & 256 &  $ 36.87\pm0.06$ \\  
2020-01-29 & 2458878.3251 & 0.943 & 237 &  $ 11.91\pm0.06$ \\ 
2020-01-30 & 2458878.5005 & 0.978 & 238 &  $  3.91\pm0.06$ \\  
2020-01-30 & 2458879.3307 & 0.144 & 277 &  $-32.46\pm0.05$ \\  
2020-01-31 & 2458879.5036 & 0.179 & 255 &  $-37.10\pm0.06$ \\  
2020-01-31 & 2458880.3289 & 0.343 & 232 &  $-34.51\pm0.06$ \\  
2020-02-01 & 2458880.5186 & 0.381 & 191 &  $-28.49\pm0.06$ \\  
\hline
\end{tabular}
\end{table}

\section{Binary system modelling}
\label{sec:binary}

We subject the TESS photometry and radial velocity measurements to simultaneous modelling to 
determine the fundamental parameters of the components of V772 Cas. To carry out this simultaneous 
modelling, we wrap the {\sc PHOEBE} binary modelling code \citep{prsa:2005,prsa:2011} into the Markov 
Chain Monte Carlo (MCMC) ensemble sampling code {\sc emcee} \citep{foreman-mackey:2013}. This 
methodology has been outlined and employed in the modelling of the magnetic systems HD~66051 
\citep{kochukhov:2018b} and HD~62658 \citep{shultz:2019} and is briefly described below. 

As discussed in detail by \citet{prsa:2005} and \citet{prsa:2016}, the {\sc PHOEBE} code employs a generalised Roche geometry model of binary systems which treats the orbital motion of eclipsing binary-star components along with the surface brightness variation caused by limb darkening, gravity darkening, reflection, ellipsoidal variations due to non-spherical shapes of the components, and brightness spots on their surfaces. The two latter phenomena are of particular interest here as possible explanations of the out-of-eclipse modulation observed in the TESS light curve of \cas. The ellipsoidal variability arises due to the changing cross-section size that faces the observer. It occurs with the orbital period and has a distinctive shape, with two maxima and minima per orbital cycle. The minimal light corresponds to eclipse phases when the components are aligned with the line of sight. In contrast, the flux variation associated with surface spots occurs with rotational periods of the components, which are not necessarily the same as the orbital period. The shapes and amplitudes of spot-induced light curves are diverse, with the brightness extrema generally not linked to particular orbital phases. In the context of {\sc PHOEBE} analysis ellipsoidal variation is an integral part of binary system modelling whereas spots are introduced with a set of additional free parameters.

\subsection{Modelling setup and results}
\label{sec:binary_results}

Given a set of input parameters {\sc PHOEBE} produces a forward binary model that consists 
of both a photometric model and radial velocity model. The aim of our modelling is to 
obtain the input parameters which produce the {\sc PHOEBE} model that best reproduces the 
entire observed TESS light curve, including eclipses and out-of-eclipse variation, and HERMES radial velocities. After we determine a 
reasonable starting model by hand, we use {\sc emcee} to explore the posterior 
distributions of the free model parameters through ensemble MCMC sampling. We use 128 
chains and initially run the code until it has reached 1000 iterations beyond convergence 
defined by a less than 1 per cent change in the estimated autocorrelation time \citep{foreman-mackey:2019}. 
After convergence is reached, we record the final positions of each chain, discard all
previous steps as burn-in, and re-initialize the algorithm from these points for an 
additional 5000 iterations. This results in 640\,000 model evaluations. We marginalize 
over all varied parameters to arrive at posterior distributions for these parameters. From 
these we draw the median and 68 per cent highest posterior density confidence intervals as the 
parameter estimate and its $1\sigma$ uncertainty. 

The TESS photometry exhibits clear eclipses with differing depths as well as out-of-eclipse variability 
on the orbital period. We fix the primary effective temperature according to the value derived in 
Sect.~\ref{sec:primary}. Although the field covered by the TESS FFI surrounding the target contains 
numerous Gaia sources, the brightest nearby target is more than five magnitudes fainter. Following this, we assumed that the light curve of \cas\ is not diluted by any additional light source and fixed
the third light parameter to zero in our {\sc PHOEBE} model. 

Preliminary binary model fits with variable eccentricity indicated that this parameter does not 
exceed 0.02 and is statistically consistent with zero. Consequently, we fixed eccentricity in the 
orbit to zero in the final analysis. This can be contrasted with a marginal eccentricity of $e=0.17\pm0.10$ inferred by \citet{gandet:2008} from low-precision photographic radial velocity measurements. Since the secondary eclipse was not detected by that study, the eccentricity could not have been constrained by the light curve.

Additionally, we apply a Gaussian prior on $v_1\sin i$, as taken 
from Sect.~\ref{sec:primary}. For each sampled parameter combination, we calculate the corresponding 
component surface gravities $\log g_1$ and $\log g_2$, and use these (along with the fixed $T_{\rm eff,1}$
and sampled $T_{\rm eff,2}$) to interpolate both limb-darkening and gravity-darkening coefficients for the 
TESS pass-band from the tables provided by \citet{claret:2017}. For limb-darkening, we apply the square-root law. 
Interpolating the gravity darkening coefficients according to the effective temperatures and surface 
gravities of the components of a given model allows us to reduce the parameter space as opposed to varying 
these coefficients and checking that they are theoretically consistent \textit{a posteriori}. 

In addition to these fixed parameters, we vary the orbital period, $P_{\rm orb}$, the reference time, $T_0$, the 
orbital inclination, $i$, the semi-major axis, $a$, the mass-ratio, $q$, the systemic velocity, $\gamma$, 
as system parameters. We also vary potentials ($\Omega_{1,2}$), albedos ($A_{1,2}$), and light contributions ($l_{1,2}$) of the primary and secondary, and vary the secondary effective temperature as well as the primary synchronicity parameter. All these parameters are allowed to vary with uniform priors, with bounds (where applicable) listed in Table~\ref{tab:binary_pars}.

\begin{table}
    \centering
    \renewcommand{\arraystretch}{1.3}
        \caption{Sampled and derived binary parameters with boundaries and estimates 
                 according to median values and uncertainties listed as 68 per cent HPD intervals.}
    
    \begin{tabular}{l c c l}
    \hline\hline
    Parameter &  & Prior range & HPD estimate \\    
    \hline
    $P_{\rm orb}$ & d & $\mathcal{U}$(0,-) & $5.0138 \pm 0.0001 $ \\
    $T_0$ & d & $\mathcal{U}$(-,-) & $2458803.4016 \pm 0.0004$ \\
    $q$ & $\frac{M_2}{M_1}$ & $\mathcal{U}$(0.01,1) &  $0.2343 \pm 0.0006$ \\
    $a$ & $R_{\odot}$ & $\mathcal{U}$(1,50) & $20.60 \pm 0.03 $ \\
    $\gamma$ & ${\rm km\,s^{-1}}$ & $\mathcal{U}$($-50$,50) & $-1.6 \pm 0.1  $ \\
    $i$ & $\deg$ & $\mathcal{U}$(45,90) & $ 85.40 \pm 0.3$ \\
    $T_{\rm eff,1}$ & ${\rm K}$ & N/A & $13\,800$ \\
    $T_{\rm eff,2}$ & ${\rm K}$ & $\mathcal{U}$(3500,50000) & $5750 \pm 150$ \\
    $A_1$ & & $\mathcal{U}$(0,1) & $0.5\pm0.3$ \\
    $A_2$ & & $\mathcal{U}$(0,1) & $0.4\pm0.03$ \\
    $\Omega_1$ &  & $\mathcal{U}$(3,20) & $4.50 \pm 0.03$\\
    $\Omega_2$ &  & $\mathcal{U}$(3,20) & $6.90 \pm 0.04$\\
    $\omega_{\rm rot,1}/\omega_{\rm orb}$ & & $\mathcal{U}$(0.1,3) & $0.459 \pm 0.007$ \\
    $l_1$ & per cent & $\mathcal{U}$(50,100) & $99.57 \pm 0.01$\\
    $l_2$ & per cent & $\mathcal{U}$(0,50) & $0.43 \pm 0.01$\\
    \hline
    $r_1/a$ & & N/A & $0.234 \pm 0.001$ \\
    $M_1$ & $M_{\odot}$ & N/A & $3.78\pm0.02$ \\
    $R_1$ & $R_{\odot}$ & N/A & $4.83\pm0.03$ \\
    $\log g_1$ & cm\,s$^{-2}$ & N/A & $3.648\pm0.005$ \\
    $r_2/a$ & & N/A & $0.0424 \pm 0.0003$ \\
    $M_2$ & $M_{\odot}$ & N/A & $0.89\pm0.01$ \\
    $R_2$ & $R_{\odot}$ & N/A & $0.87\pm0.01$ \\
    $\log g_2$ & cm\,s$^{-2}$ & N/A & $4.50\pm0.01$ \\

    \hline
    \end{tabular}
    \label{tab:binary_pars}
\end{table}

\begin{figure}
\centering
\includegraphics[width=0.95\columnwidth]{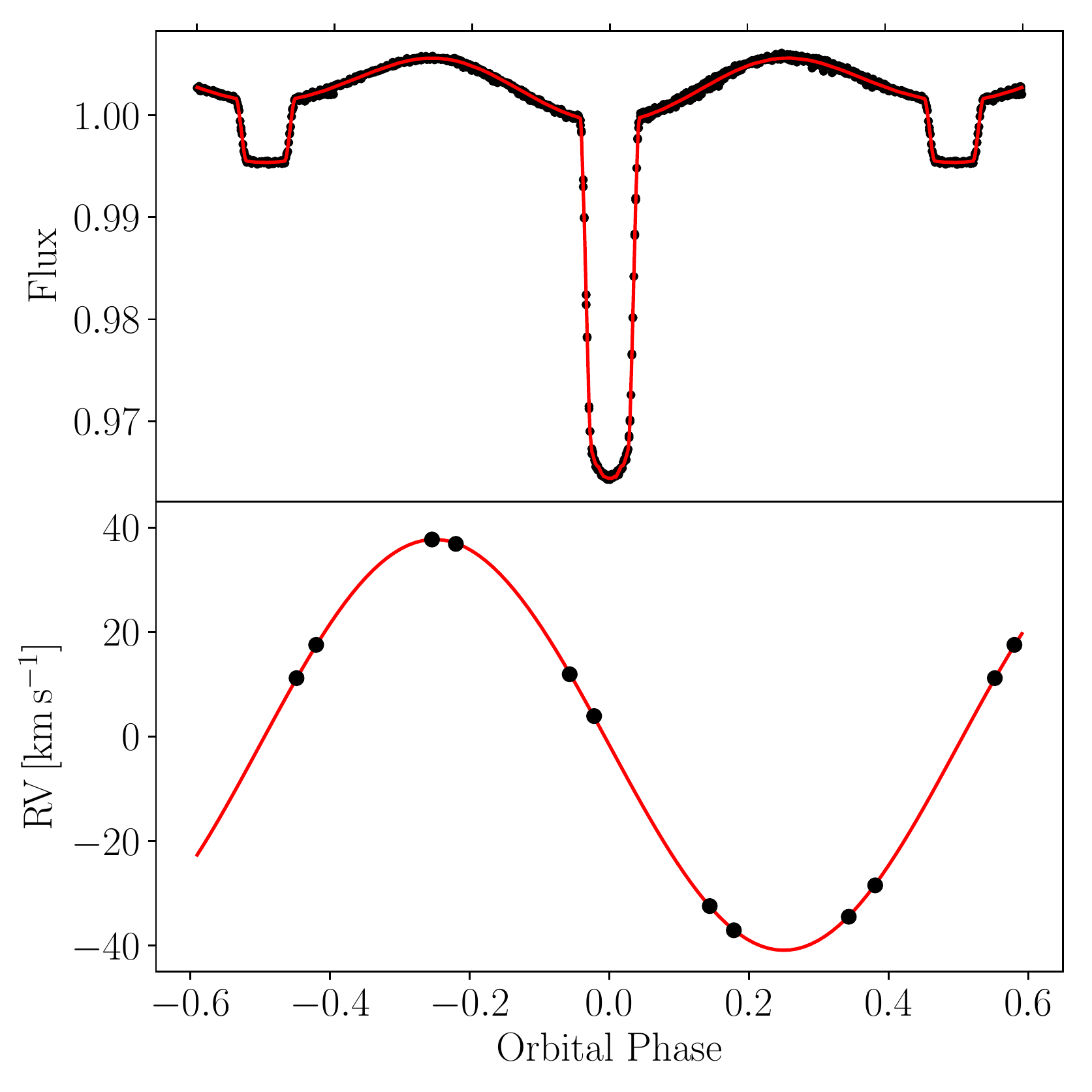}
\caption{{\bf Top}: TESS FFI observations (black symbols) and optimized {\sc PHOEBE} light curve model (red line) as a function of the orbital phase. {\bf Bottom}: Radial velocities derived from HERMES spectra (black symbols) and optimized {\sc PHOEBE} radial velocity model (red line).}
\label{fig:eb_soln}
\end{figure}

\begin{figure}
\centering
\includegraphics[width=\columnwidth]{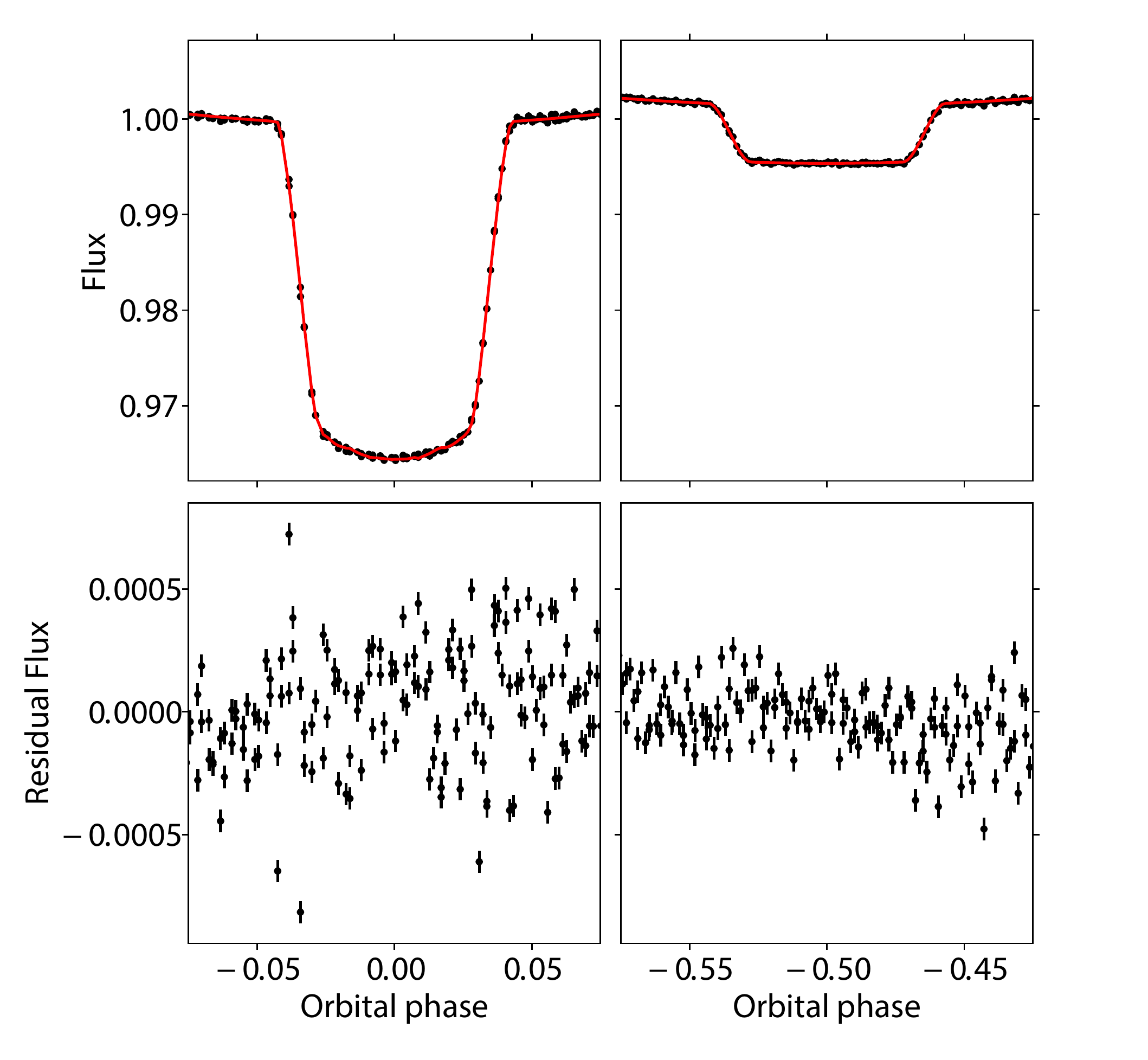}
\caption{{\bf Top}: TESS FFI observations (black symbols) and optimized {\sc PHOEBE} light curve model (red line) in the vicinity of primary (left) and secondary (right) eclipses. {\bf Bottom}: Residuals from the light curve fit.}
\label{fig:eclp}
\end{figure}

\begin{figure*}
\centering
\figps{\hsize}{90}{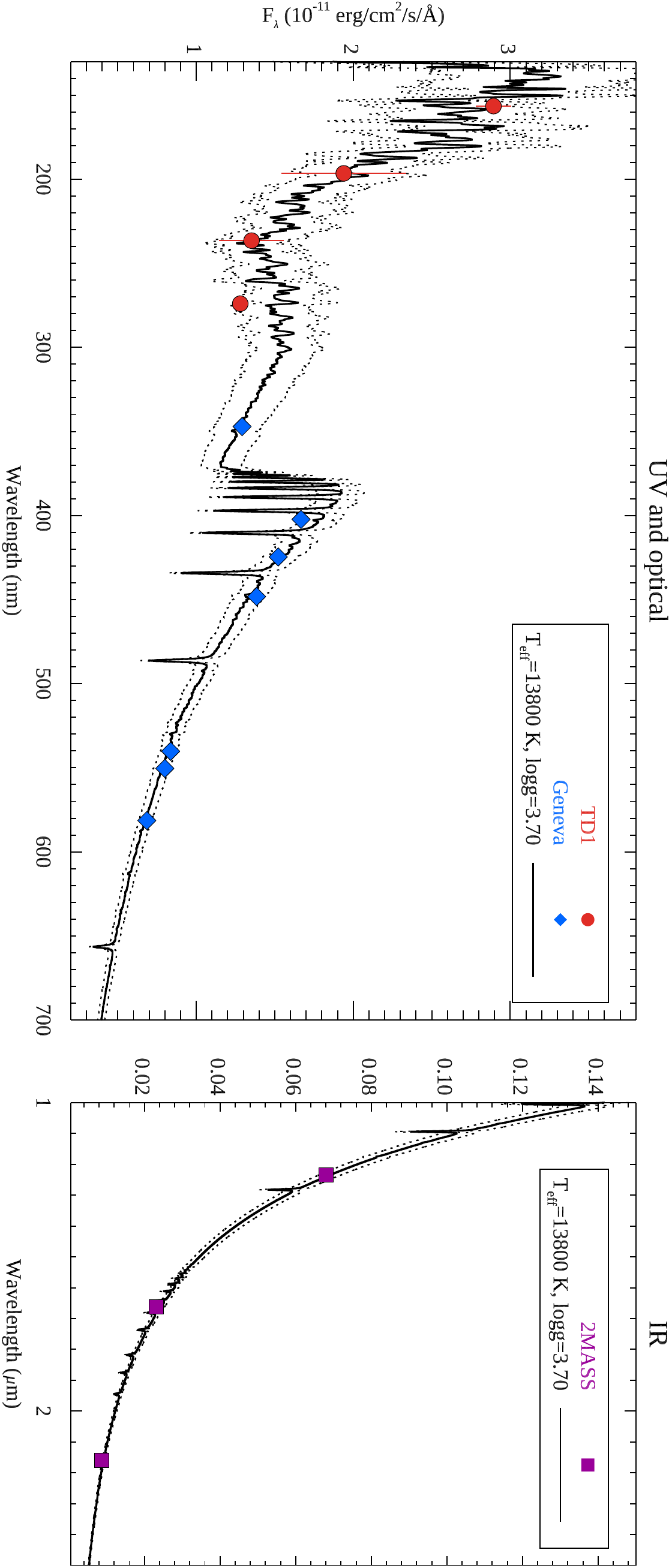}
\caption{Comparison between the observed (symbols) and theoretical (lines) spectral energy distribution of \cas\ in the ultraviolet and optical (left panel) and near-infrared (right panel) wavelength regions. Calculation shown with the thick solid line corresponds to primary's parameters \teff\,=\,13800~K, \logg\,=\,3.7, $R=4.95$~\rsun, and reddening $E(B-V)=0.144$. Dotted lines illustrate the impact of changing \teff\ by $\pm500$~K.}
\label{fig:sed}
\end{figure*}

Despite the lack of detection of the secondary in the spectra or LSD profiles, the presence of 
flat bottom secondary eclipse (which indicates a total eclipse) enables us to obtain some extra constraints on the stellar radii, and through the 
determination of the potentials, the mass ratio. Additionally, the presence of ellipsoidal 
variability provides further constraint on the mass ratio. The optimised parameter estimates and
their highest posterior density (HPD) 68 per cent ($1\sigma$) uncertainties are listed in Table~\ref{tab:binary_pars}. 
The posterior distributions are illustrated in Figs.~\ref{fig:pd1} to \ref{fig:pd4}.
Our solution reports an evolving
intermediate mass $M_1=3.78\pm0.02$~$M_{\odot}$, $R_1=4.83\pm0.03$~$R_{\odot}$ primary and 
a lower mass $M_2=0.89\pm0.01$~$M_{\odot}$, $R_2=0.87\pm0.01$~$R_{\odot}$ secondary which is 
still in the first half of its main-sequence evolution. The size of both components is well below their Roche radii (10.8~$R_{\odot}$ for the primary and 5.0~$R_{\odot}$ for the secondary, respectively).

The optimized light curve and radial velocity models constructed from these parameters are shown in the top and bottom panels of Fig.~\ref{fig:eb_soln}, respectively. This figure demonstrates that the PHOEBE binary system model successfully reproduces available photometric and spectroscopic observations, both within the primary and secondary eclipses and outside the eclipses. The latter 0.6 per cent peak-to-peak photometric variation is thus interpreted as an ellipsoidal variability caused by a slightly distorted shape of the primary. Our analysis indicates that its maximum deviation from a spherical shape is about 0.7 per cent. In addition, Fig.~\ref{fig:eclp} shows observations, the light curve model, and the corresponding residuals around the primary and secondary eclipses. The residuals shown in the lower panel of Fig.~\ref{fig:eclp} reveal no systematic trends.

We do not include spots
in our binary model as adding those would introduce several additional free parameters. There is, in fact, no evidence of spot signatures in the photometric data after ellipsoidal variation is accounted for. Inspection of the residuals after subtraction of the binary light curve model does not reveal any clear indication of rotational modulation or any other periodic variability above $\approx100$~ppm for the frequencies below 2 d$^{-1}$ and 20~ppm in the 2--24 d$^{-1}$ frequency range. 

Given the sharp points of ingress and egress (the first and fourth contacts) at both primary and secondary eclipse, we ran two {\sc PHOEBE} models with different fine and coarse grid sizes to characterise the numerical noise in our solution. We find a difference of 18~ppm between the two models, whereas the residual scatter is 230~ppm, meaning that the numerical noise is responsible for less than 10 per cent of the residual scatter \citep{maxted:2020}.

\begin{figure}
\centering
\figps{0.85\hsize}{0}{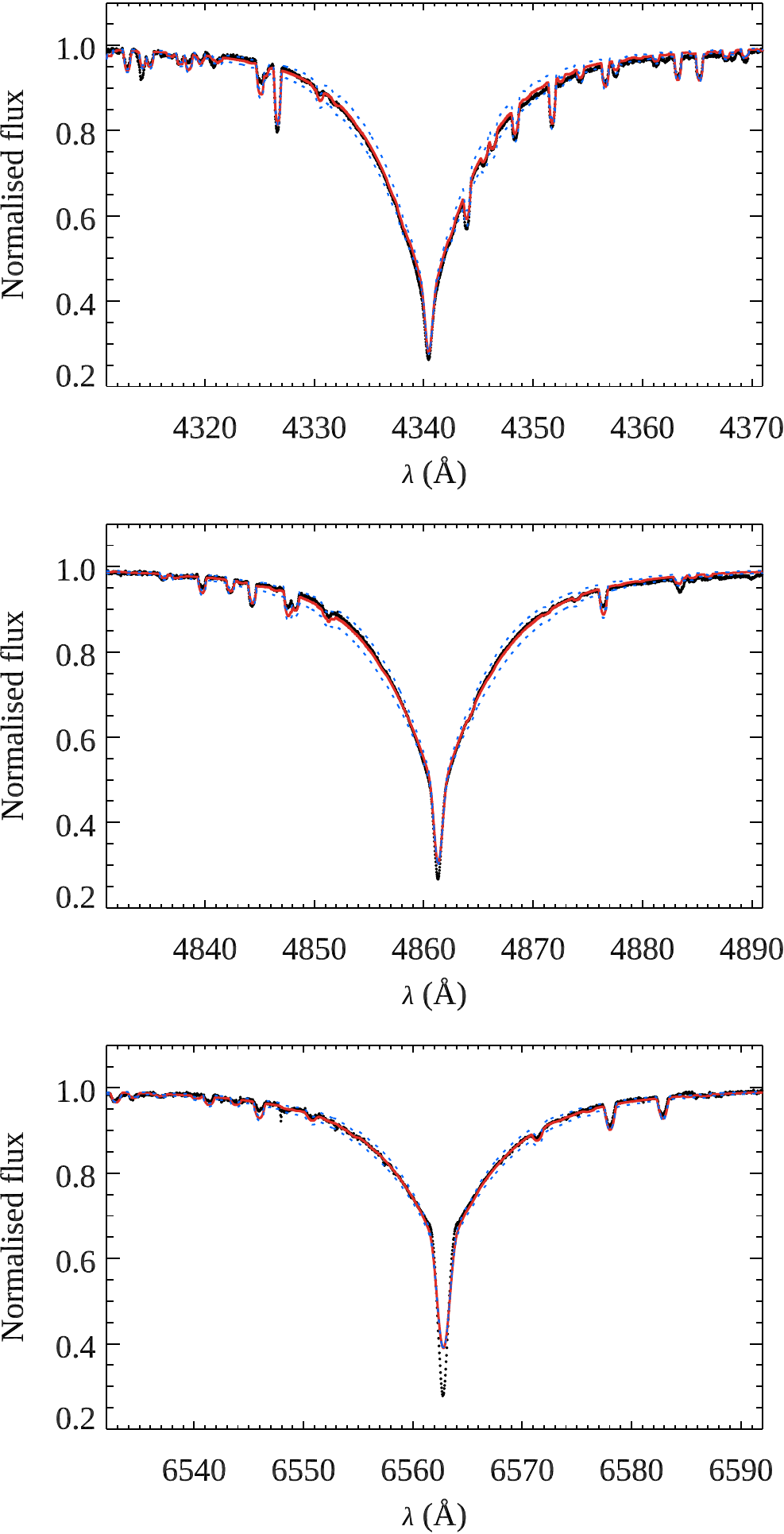}
\caption{Comparison between the observed (symbols) and theoretical (lines) hydrogen Balmer line profiles. Calculations shown with solid lines employ \teff\,=\,13800~K, \logg\,=\,3.7 model atmosphere for the primary together with the chemical abundances determined in this study. Dotted lines illustrate the effect of changing \logg\ by $\pm0.2$~dex.}
\label{fig:hlines}
\end{figure}

\begin{figure*}
\centering
\figps{\hsize}{0}{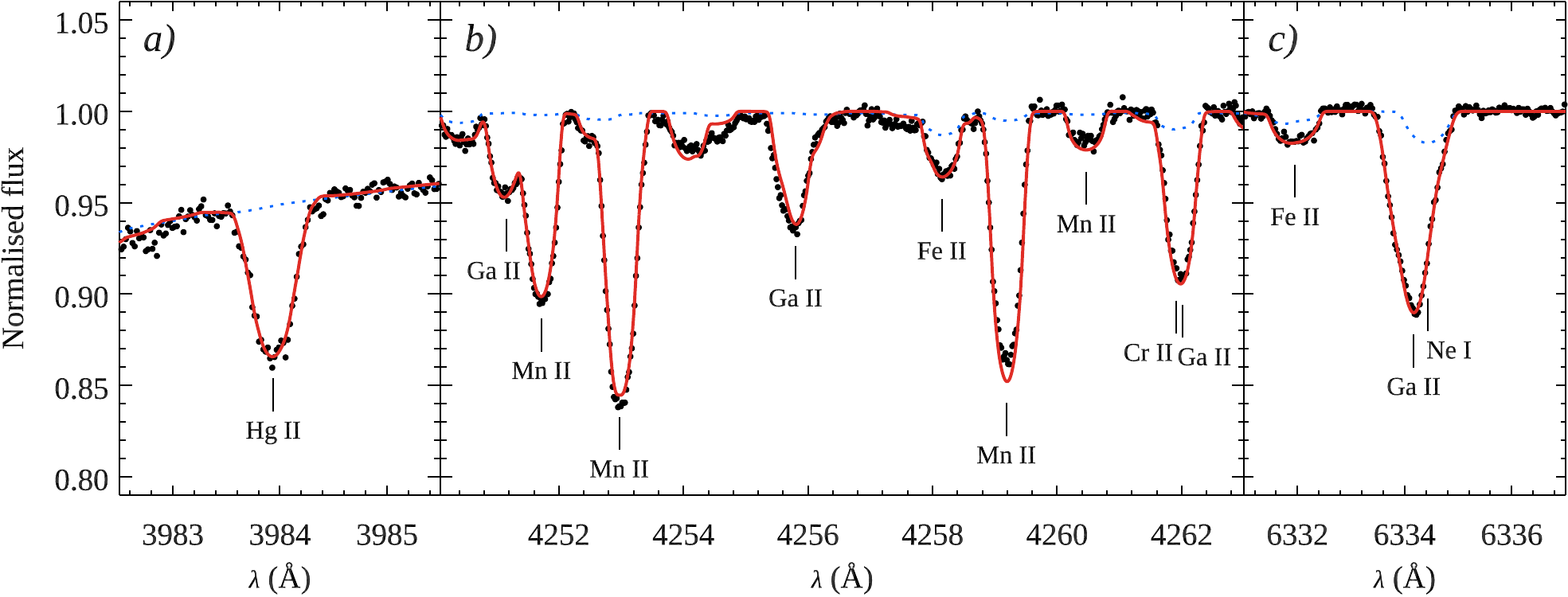}
\caption{Comparison of the average spectrum of \cas\ (symbols) with the best fitting theoretical model spectrum (solid line) in several wavelength regions containing spectral lines commonly enhanced in HgMn stars. Dotted line shows synthetic spectrum calculated with solar abundances.}
\label{fig:fits}
\end{figure*}

\begin{figure}
\centering
\figps{\hsize}{0}{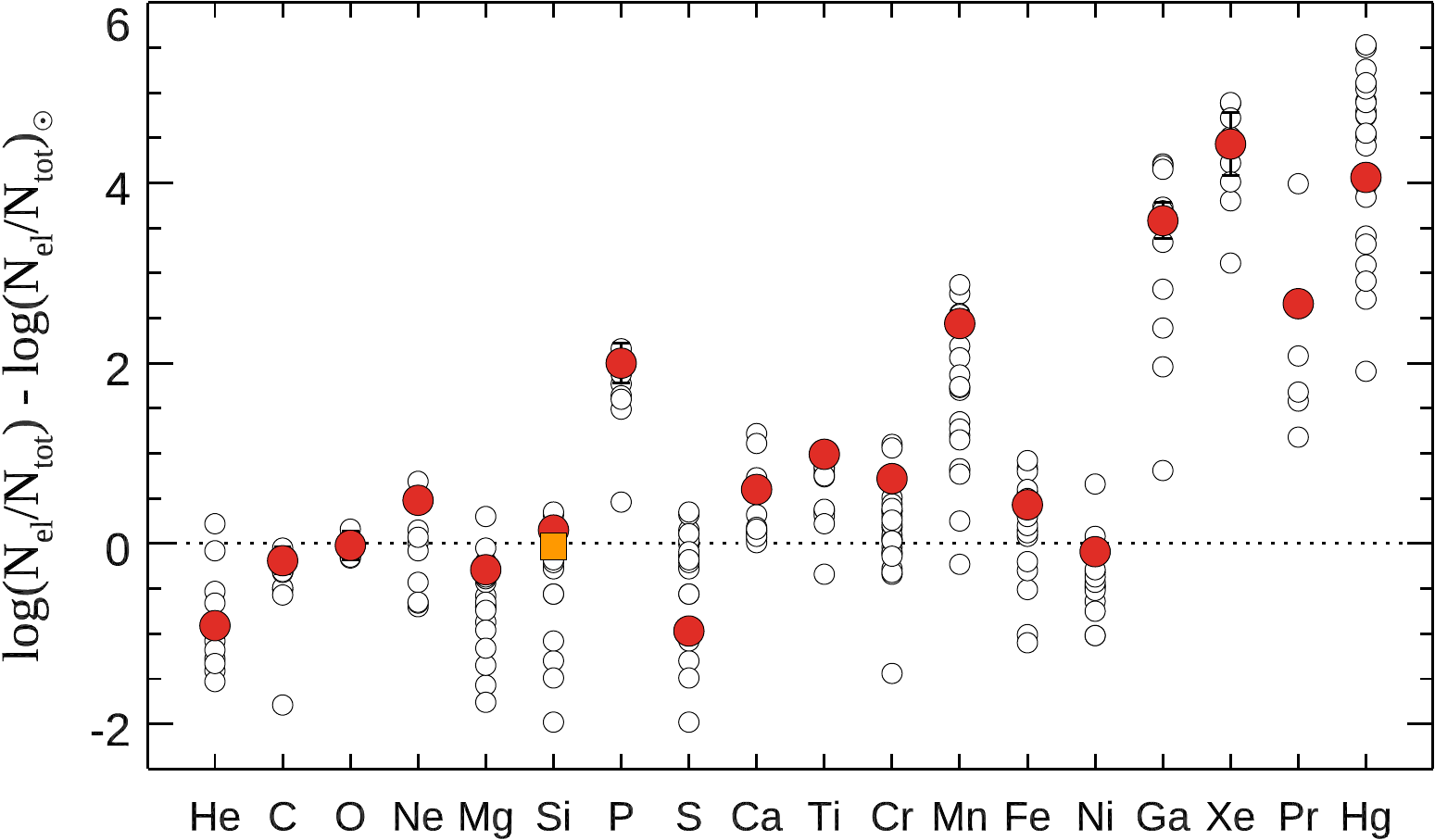}
\caption{Abundances of individual elements in the primary component of \cas\ (filled symbols) relative to the solar chemical composition. Filled circles correspond to neutral and singly ionised species; the filled square shows the abundance of \ion{Si}{iii}. Open symbols illustrate abundances of HgMn stars with \teff\,=\,13000--14000~K from the catalogue by \citet{ghazaryan:2018}.}
\label{fig:abund}
\end{figure}

\section{Analysis of HgMn primary}
\label{sec:primary}

\subsection{Atmospheric parameters}

Here we use a combination of modelling the stellar spectral energy distribution (SED) and the hydrogen Balmer lines to determine primary's \teff\ and \logg\ respectively. This approach is commonly employed for normal and peculiar late-B and A-type stars (e.g. \citealt*{ryabchikova:1999a}; \citealt{fossati:2009}; \citealt{rusomarov:2016}; \citealt*{kochukhov:2019}). Other spectral indicators, such as \ion{He}{i} and metal lines, cannot be used for the atmospheric parameter determination of HgMn stars due to non-solar photospheric element abundances and occasional vertical chemical stratification.

Model atmosphere analysis of the primary component of \cas\ was carried out using the {\sc LLmodels} code \citep{shulyak:2004}, taking into account individual atmospheric abundances. The influence of the faint secondary was ignored. The effective temperature was determined by comparing the model SED with the TD1 satellite stellar flux measurements in UV \citep{thompson:1978} as well as the optical and near-infrared fluxes obtained from Geneva \citep{hauck:1982} and 2MASS \citep{cutri:2003} photometric measurements, respectively. The reddening $E(B-V)=0.144\pm0.050$ \citep{lallement:2019} and the Gaia DR2 distance $363\pm6$~pc \citep{gaia-collaboration:2018} were adopted for SED fitting with the stellar effective temperature and radius adjusted to match the observations. This analysis yielded \teff\,=\,$13800\pm500$~K and $R=4.95\pm0.2$~\rsun. Figure~\ref{fig:sed} compares observations with the theoretical SED computed with {\sc LLmodels} using these parameters. The stellar effective temperature derived from the SED agrees reasonably well with \teff\,=\,13400--13800~K that can be obtained for this star using different Str\"omgren and Geneva photometric calibrations (\citealt{kunzli:1997}; \citealt*{paunzen:2005a}; \citealt*{paunzen:2006}).

The surface gravity was determined by fitting the observed hydrogen Balmer line profiles in the time-averaged HERMES spectrum with the theoretical calculations using the {\sc Synth3} spectrum synthesis code \citep{kochukhov:2007d} and the {\sc LLmodels} atmospheres described above. The average spectrum was constructed by co-adding ten individual observations after correcting the radial velocity shifts reported in Table~\ref{tab:obs}. As demonstrated by Fig.~\ref{fig:hlines}, the hydrogen lines in the mean spectrum of the primary are well approximated with \logg\,=\,$3.7\pm0.1$. This spectroscopic estimate of \logg\ and the stellar radius inferred from the SED are consistent within uncertainty with the results of binary system modelling with {\sc PHOEBE} in Sect.~\ref{sec:binary_results}.

\subsection{Abundances}

Chemical abundances were estimated by fitting {\sc Synth3} spectra to short segments of the average spectrum of \cas. In these fits, individual abundances of one or several elements as well as the projected rotational velocity were determined using the {\sc BinMag} \citep{kochukhov:2018}\footnote{\url{https://www.astro.uu.se/~oleg/binmag.html}} IDL tool. Different lines of the same ions were analysed independently and the scatter of abundances, quantified by the standard deviation, was taken as an uncertainty estimate. The input line list for these calculations was extracted from VALD, using the latest version of the database that incorporates hyperfine and isotopic splitting \citep{pakhomov:2019}. For \cas\ the former is particularly important for accurate analysis of \ion{Mn}{ii} and \ion{Ga}{ii} lines. Information on the hyperfine and isotopic splitting of the \ion{Hg}{ii} 3984~\AA\ line was taken from \citet{woolf:1999}.

Several examples of spectrum synthesis fits around the lines typically enhanced in HgMn stars are shown in Fig.~\ref{fig:fits}. Examples of fits to other ions analysed in the paper are demonstrated in Fig.~\ref{fig:other_fits}. We were able to obtain abundance estimates for 19 ions based on 164 individual lines and blends, listed in Table~\ref{tab:lines}. The resulting abundances are presented in Table~\ref{tab:abund}, which lists the ions studied, the number of lines analysed, the average abundance and corresponding error for \cas, the corresponding average abundance of HgMn stars with \teff\,=\,13000--14000~K from the compilation by \citet{ghazaryan:2018}, and the solar abundance of each ion \citep{asplund:2009}. Our abundance analysis of \cas\ was carried out under the local thermodynamic equilibrium (LTE) assumption. Several abundances reported in Table~\ref{tab:abund}, for example Ca \citep*{sitnova:2018}, Si \citep{mashonkina:2020}, and Ne \citep{alexeeva:2020}, are likely affected by departures from LTE leading to abundance corrections of about 0.1--0.3~dex.

The abundance pattern of \cas\ relative to the solar chemical composition is illustrated in Fig.~\ref{fig:abund}. We also show in this figure abundances of 33 HgMn stars in the 13000--14000~K \teff\ range from the catalogue by \citet{ghazaryan:2018}. It is evident that \cas\ exhibits an unremarkable abundance pattern, very often seen in HgMn stars with a similar \teff. In particular, He is underabundant, Si is close to solar, P and Mn are overabundant by $\sim$\,2~dex, and Ga, Xe, and Hg are overabundant by up to 4~dex. Pr is also overabundant in \cas\ by $\approx$\,2.5~dex, as found for several other HgMn stars, although its abundance estimate based on a single blended \ion{Pr}{iii} line is somewhat uncertain.

Considering the projected rotational velocity, determined with the spectrum synthesis fit from 84 unblended lines, we found \vsini\,=\,$22.3\pm0.3$~\kms. Together with the stellar radius $R=4.83\pm0.03$~\rsun\ and inclination of the rotation axis $i_{\rm rot}=i_{\rm orb}=85.4\pm0.3$\degr\ found in Sect.~\ref{sec:binary}, this implies $P_{\rm rot}=10.9\pm0.2$~d. Thus, the primary is rotating sub-synchronously with a period about twice longer than the orbital period provided that the orbital and rotational axes are aligned. The latter assumption is reasonable for close binaries on theoretical grounds \citep{zahn:1977,hut:1981} and generally agrees with observational findings \citep{hale:1994,farbiash:2004}.

\begin{table}
\caption{Atmospheric chemical composition of the primary component of \cas. The columns give the ion identification, the number of lines studied, abundance for \cas, average abundance for HgMn stars in the 13000--14000 \teff\ interval \citep{ghazaryan:2018}, and the solar abundance \citep{asplund:2009}. \label{tab:abund}}
 \begin{tabular}{lllll}
\hline\hline
Ion & $N$ & \multicolumn{3}{c}{$\log(N_{\rm el}/N_{\rm tot})$} \\
\cline{3-5} \\[-0.2cm]
& & \cas\ & $\langle \mathrm{HgMn} \rangle$ & Sun \\
\hline
\ion{He}{i}   &   5 & $-2.02\pm0.10$  & $-2.01\pm0.52$ & $ -1.11$ \\
\ion{C}{ii}   &   3 & $-3.80\pm0.15$  & $-4.05\pm0.48$ & $ -3.61$ \\
\ion{O}{i}    &   2 & $-3.37\pm0.16$  & $-3.36\pm0.13$ & $ -3.35$ \\
\ion{Ne}{i}   &  11 & $-3.63\pm0.04$  & $-4.29\pm0.47$ & $ -4.11$ \\
\ion{Mg}{ii}  &   3 & $-4.73\pm0.14$  & $-5.16\pm0.53$ & $ -4.44$ \\
\ion{Si}{ii}  &   5 & $-4.38\pm0.09$  & $-4.89\pm0.63$ & $ -4.53$ \\
\ion{Si}{iii} &   2 & $-4.56\pm0.01$  & $-4.89\pm0.63$ & $ -4.53$ \\
\ion{P}{ii}   &  21 & $-4.63\pm0.22$  & $-4.94\pm0.48$ & $ -6.63$ \\
\ion{S}{ii}   &   8 & $-5.89\pm0.10$  & $-5.28\pm0.63$ & $ -4.92$ \\
\ion{Ca}{ii}  &   1 & $-5.10       $  & $-5.32\pm0.41$ & $ -5.70$ \\
\ion{Ti}{ii}  &   8 & $-6.10\pm0.07$  & $-6.58\pm0.36$ & $ -7.09$ \\
\ion{Cr}{ii}  &   6 & $-5.68\pm0.06$  & $-6.24\pm0.51$ & $ -6.40$ \\
\ion{Mn}{ii}  &  38 & $-4.17\pm0.14$  & $-4.97\pm0.86$ & $ -6.61$ \\
\ion{Fe}{ii}  &  42 & $-4.11\pm0.10$  & $-4.32\pm0.50$ & $ -4.54$ \\
\ion{Ni}{ii}  &   1 & $-5.91       $  & $-6.20\pm0.43$ & $ -5.82$ \\
\ion{Ga}{ii}  &   3 & $-5.42\pm0.20$  & $-5.87\pm1.13$ & $ -9.00$ \\
\ion{Xe}{ii}  &   3 & $-5.37\pm0.35$  & $-5.51\pm0.58$ & $ -9.80$ \\
\ion{Pr}{iii} &   1 & $-8.66       $: & $-9.21\pm1.10$ & $-11.32$ \\ 
\ion{Hg}{ii}  &   1 & $-6.81       $  & $-6.49\pm1.09$ & $-10.87$ \\
\hline
\end{tabular}
\end{table}

\section{Discussion}
\label{sec:disc}

In this paper we investigated the nature of the bright but poorly studied eclipsing binary system \cas. Based on the new high-precision photometric data provided by the TESS mission we confirmed the presence of primary eclipses and identified secondary eclipses for the first time. Significant out-of-eclipse photometric variability, synchronised with the orbital motion, was confirmed for this system. We have acquired high-resolution spectra of \cas\ with the aim to determine fundamental parameters of the components and better characterise surface abundance pattern of the primary star, which in the past was attributed conflicting HgMn and BpSi spectral classifications. 

Our analysis reveals that \cas\ is an SB1 system with a late-B primary showing abundance anomalies typical of HgMn stars. Measurement of the projected rotational velocity shows that the primary rotates sub-synchronously. Taking into account constraints from the atmospheric modelling of the primary, we carried out detailed modelling of the TESS light curve with the PHOEBE code. This analysis demonstrated that the conspicuous out-of-eclipse variability is explained by the ellipsoidal shape of the primary, which arises due to the tidal interaction with the secondary. The ellipsoidal variability is clearly visible in the TESS light curve despite a relatively low mass of the secondary thanks to a high precision of the space photometry data. Interpretation of the ellipsoidal variability together with eclipses and radial velocity variation of the primary allowed us to retrieve accurate masses and radii of both components despite the absence of the secondary's lines in the optical spectra. We concluded that an evolved HgMn primary of \cas\ is orbited by a main sequence solar-type secondary, resulting in large luminosity and mass ratios.

Our results confirm that \cas\ should be formally classified as an Algol-type (EA) eclipsing binary \citep{samus:2017}, without any additional types of variability present. In particular, the absence of rotational modulation due to surface spots and the HgMn spectroscopic classification of the primary demonstrates that this star is not an $\alpha^2$~CVn-type magnetic variable similar to the magnetic Bp stars recently identified in the eclipsing binaries HD~66051 \citep{kochukhov:2018b} and HD~65658 \citep{shultz:2019}. Instead, \cas\ should be discussed in the context of research on HgMn stars. These objects lack strong large-scale magnetic fields \citep{shorlin:2002,auriere:2010a,bagnulo:2012,kochukhov:2011b,makaganiuk:2011a,makaganiuk:2011,makaganiuk:2012} and also possess no complex tangled fields stronger than a few hundred G \citep{kochukhov:2013a}. Despite this, these stars are able to develop a low-contrast non-uniform surface abundance distribution of some heavy elements \citep{adelman:2002,kochukhov:2005,folsom:2010,makaganiuk:2011}. Geometry of these abundance spots appears to slowly evolve with time \citep{kochukhov:2007b,briquet:2010,korhonen:2013}. Photometric time series studies of HgMn stars occasionally reveal rotational modulation, presumably related to heavy-element abundance spots (\citealt{morel:2014}; \citealt{strassmeier:2017}; \citealt{white:2017}; \citealt*{prvak:2020}), and, possibly, to SPB pulsations \citep{hummerich:2018}. Our study shows that ellipsoidal variability is yet another phenomenon that can contribute to or even dominate the light curves of HgMn stars in close binaries.

\begin{table*}
\caption{Properties of eclipsing binary systems containing HgMn stars. When available, uncertainties in the last significant digit are indicated by the numbers in brackets. \label{tab:hgmns}}
 \begin{tabular}{lllllllllll}
\hline
\hline
Star & $V$ & Spec & $P_{\rm orb}$ (d) & $e$ & $q$ & $M_1$ ($M_\odot$) & $R_1$ ($R_\odot$) & $M_2$ ($M_\odot$) & $R_2$ ($R_\odot$) & Reference \\
\hline                                  
AR Aur          &   6.14 & SB2 &    4.13 & 0.0    & 0.927(2) & 2.544(9) & 1.80(1) & 2.358(8) & 1.83(2) &  \citet{hubrig:2012} \\
TYC 455-791-1   &  11.95 & SB2 &   12.47 & 0.18   & 0.941(8)  & 3.1  & 2.4  & 2.9  & 2.3  &  \citet{strassmeier:2017} \\
V772 Cas        &   6.68 & SB1 &    5.01 & 0.0    & 0.2343(6) & 3.78(2) & 4.83(3) & 0.89(1) & 0.87(1) &  This work \\
\hline
\end{tabular}
\end{table*}

\begin{figure}
\centering
\figps{\hsize}{0}{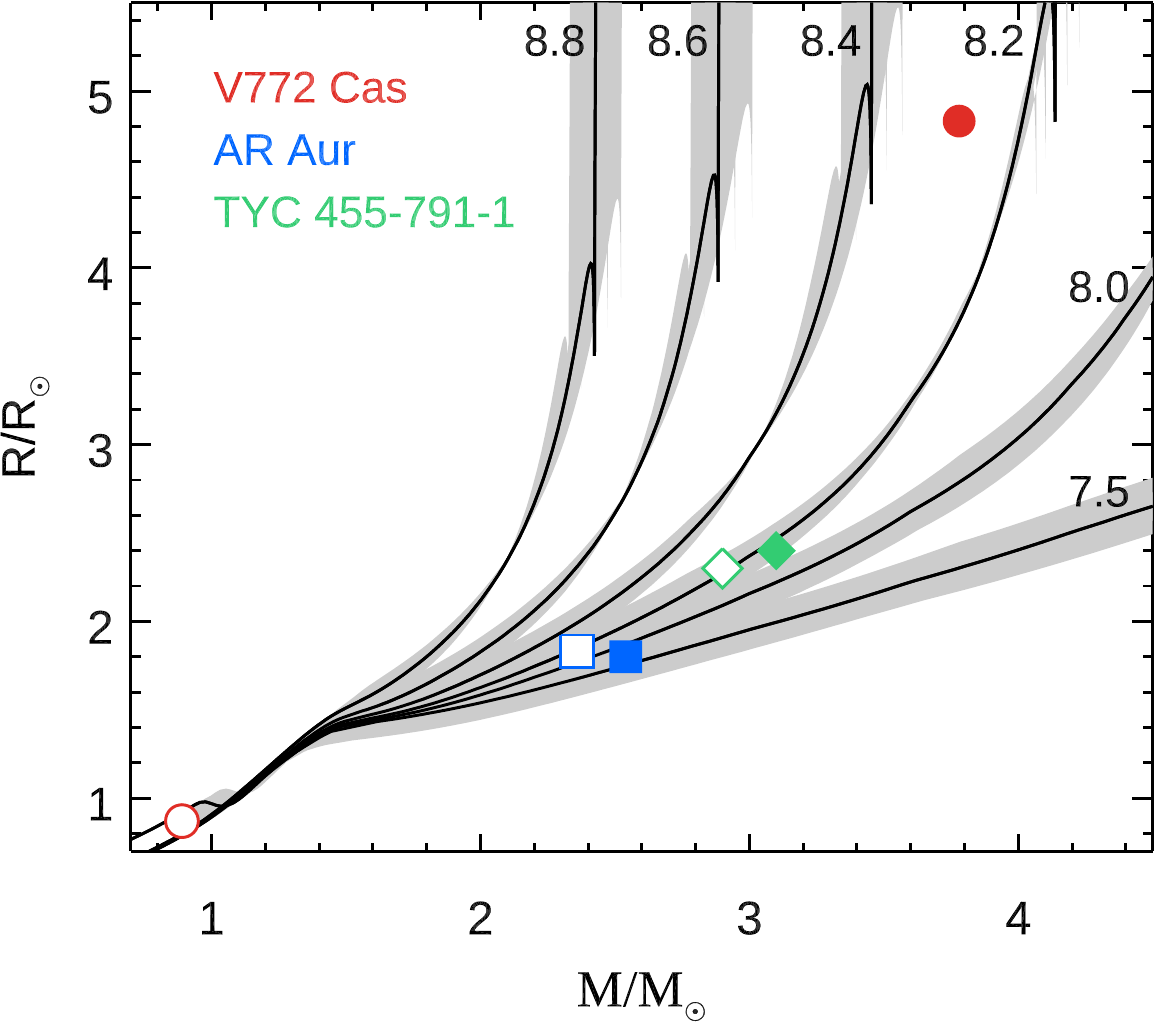}
\caption{Masses and radii of the eclipsing binaries with HgMn components. Parameters of the primaries are indicated by filled symbols. Open symbols correspond to the secondaries. The three systems included in this plot are \cas\ (circles), AR~Aur (squares), and TYC 455-791-1 (diamonds). Theoretical MESA isochrones corresponding to ages from $\log t/\mathrm{yr}=7.5$ to 8.8 are also shown. The underlying grey curves illustrate the effect of varying metallicity by $\pm0.2$~dex.}
\label{fig:isoc}
\end{figure}

Although HgMn stars are commonly found in close binary systems \citep{abt:1973,gerbaldi:1985}, only two eclipsing binaries containing HgMn components were known prior to this study. They are AR~Aur \citep{nordstrom:1994,folsom:2010,hubrig:2012} and TYC 455-791-1 \citep{strassmeier:2017}. Table~\ref{tab:hgmns} summarises the properties of these three systems in comparison to \cas. All three eclipsing binaries are non-interacting, detached systems, so their components should have evolved as if they were single stars (\citealt*{torres:2010}). The resulting mass-radius relationship is compared to single-star MESA (Modules for Experiments in Stellar Astrophysics) theoretical stellar evolution models from the MIST grid\footnote{\url{http://waps.cfa.harvard.edu/MIST}} \citep{dotter:2016,choi:2016} in Fig.~\ref{fig:isoc}. These models are available for two values of initial rotational velocity, a wide metallicity range, and a single empirically calibrated overshooting prescription, consistent with recent observational constraints \citep{claret:2019}. In this work, we have chosen to use the isochrones that were computed ignoring the stellar rotation. This assumption is appropriate for the three systems considered here and for HgMn stars in general since a slow rotation is known to be a necessary condition for this chemical peculiarity to appear \citep{michaud:1982}.

The chemical anomalies produced by the atomic diffusion in A and B stars are constrained to the outermost stellar layers and are not indicative of the bulk metallicity (e.g. \citealt*{richard:2001}). The latter is unknown but is believed to not differ much from that of normal stars. According to \citet{sofia:2001}, the metallicity scatter of young F and G disk stars in the solar neighbourhood reaches 0.1--0.2~dex, which is similar to the 0.4~dex metallicity range of open clusters containing ApBp stars \citep{bagnulo:2006,landstreet:2007}. Therefore, we included a metallicity variation by $\pm0.2$~dex around the solar value when estimating the age of \cas\ from the MIST isochrones.

In all three systems the star with HgMn peculiarity is the primary. AR~Aur and TYC 455-791-1 are close to ZAMS and have mass ratios not far from unity. In fact, AR~Aur~B is probably still contracting towards ZAMS, indicating the extreme youth of this system \citep{nordstrom:1994}. The masses of the components of AR~Aur and TYC 455-791-1 span a narrow range of 2.4--3.1 \msun. On the other hand, the primary of \cas\ is more massive and considerably more evolved. It is likely to be at or near the TAMS whereas the solar-type secondary is still on the main sequence. The location of \cas\ components on the mass-radius diagram illustrated in Fig.~\ref{fig:isoc} is best described by a set of isochrones with ages from $\log t/\mathrm{yr}=8.26$ to 8.32 for the metallicity range discussed above and the standard overshooting prescription adopted in the MIST grid. A similar age range (8.32--8.38) is obtained by applying the isochrone-cloud fitting methodology \citep{johnston:2019} with a different grid of solar-metallicity MESA models spanning a large range of overshooting parameter.

Despite this dramatic difference in evolutionary status, the HgMn primaries of all three eclipsing binary systems share the same basic chemical abundance pattern. This observation suggests that the HgMn-type chemical peculiarity develops rapidly at or near the ZAMS and then persists throughout the entire main sequence life time of an intermediate-mass star. This is broadly in agreement with the investigation of H-R diagram positions of single HgMn stars by \citet*{adelman:2003a}, although that study showed that these stars tend to concentrate in the first half of the main-sequence life time and that there are very few luminous HgMn stars comparable to \cas\ A.

\section*{Acknowledgements}
Based on observations made with the Mercator Telescope, operated on the Island of La Palma by the Flemish Community, at the Spanish Observatorio del Roque de los Muchachos of the Instituto de Astrof\'isica de Canarias.
This work has made use of the VALD database, operated at Uppsala University, the Institute of Astronomy RAS in Moscow, and the University of Vienna. This research has made use of the SIMBAD database, operated at CDS, Strasbourg, France. Some of the data presented in this paper were obtained from the Mikulski Archive for Space Telescopes (MAST). 
O.K. acknowledges support by the Swedish Research Council and the Swedish National Space Board.
J.L.-B. acknowledges support from FAPESP (grant 2017/23731-1).
This paper includes data collected by the TESS mission, which are publicly available from the Mikulski Archive for Space Telescopes (MAST). Funding for the TESS mission is provided by NASA's Science Mission directorate. 
This research made use of Astropy,\footnote{\url{http://www.astropy.org}} a community-developed core Python package for Astronomy \citep{astropy:2013,astropy:2018}
as well as the Corner\footnote{\url{https://corner.readthedocs.io}} Python code by \citet{foreman-mackey:2016}.
The research leading to these results has received funding from the European 
Research Council (ERC) under the European Union's Horizon 2020 research and 
innovation programme (grant agreement N$^\circ$670519: MAMSIE), from the 
KU\,Leuven Research Council (grant C16/18/005: PARADISE), from the Research 
Foundation Flanders (FWO) under grant agreements G0H5416N (ERC Runner Up Project) 
and G0A2917N (BlackGEM), as well as from the BELgian federal Science Policy Office 
(BELSPO) through PRODEX grant PLATO.
T.R. thanks the Ministry of Science and Higher Education of Russian Federation
(grant 13.1902.21.0039) for partial financial support.
M.E.S. acknowledges  support from the Annie Jump Cannon Fellowship, supported by the University of Delaware and endowed by the Mount Cuba Astronomical Observatory.

\section*{Data availability}
The data underlying this article will be shared on reasonable request to the corresponding author.


\begin{thebibliography}{}
\makeatletter
\relax
\def\mn@urlcharsother{\let\do\@makeother \do\$\do\&\do\#\do\^\do\_\do\%\do\~}
\def\mn@doi{\begingroup\mn@urlcharsother \@ifnextchar [ {\mn@doi@}
  {\mn@doi@[]}}
\def\mn@doi@[#1]#2{\def\@tempa{#1}\ifx\@tempa\@empty \href
  {http://dx.doi.org/#2} {doi:#2}\else \href {http://dx.doi.org/#2} {#1}\fi
  \endgroup}
\def\mn@eprint#1#2{\mn@eprint@#1:#2::\@nil}
\def\mn@eprint@arXiv#1{\href {http://arxiv.org/abs/#1} {{\tt arXiv:#1}}}
\def\mn@eprint@dblp#1{\href {http://dblp.uni-trier.de/rec/bibtex/#1.xml}
  {dblp:#1}}
\def\mn@eprint@#1:#2:#3:#4\@nil{\def\@tempa {#1}\def\@tempb {#2}\def\@tempc
  {#3}\ifx \@tempc \@empty \let \@tempc \@tempb \let \@tempb \@tempa \fi \ifx
  \@tempb \@empty \def\@tempb {arXiv}\fi \@ifundefined
  {mn@eprint@\@tempb}{\@tempb:\@tempc}{\expandafter \expandafter \csname
  mn@eprint@\@tempb\endcsname \expandafter{\@tempc}}}

\bibitem[\protect\citeauthoryear{{Abt} \& {Snowden}}{{Abt} \&
  {Snowden}}{1973}]{abt:1973}
{Abt} H.~A.,  {Snowden} M.~S.,  1973, \apjs, \href
  {http://adsabs.harvard.edu/abs/1973ApJS...25..137A} {25, 137}

\bibitem[\protect\citeauthoryear{{Adelman}, {Gulliver}, {Kochukhov}  \&
  {Ryabchikova}}{{Adelman} et~al.}{2002}]{adelman:2002}
{Adelman} S.~J.,  {Gulliver} A.~F.,  {Kochukhov} O.~P.,   {Ryabchikova} T.~A.,
  2002, \mn@doi [\apj] {10.1086/341140}, \href
  {http://adsabs.harvard.edu/abs/2002ApJ...575..449A} {575, 449}

\bibitem[\protect\citeauthoryear{{Adelman}, {Adelman}  \& {Pintado}}{{Adelman}
  et~al.}{2003}]{adelman:2003a}
{Adelman} S.~J.,  {Adelman} A.~S.,   {Pintado} O.~I.,  2003, \mn@doi [\aap]
  {10.1051/0004-6361:20021475}, \href
  {https://ui.adsabs.harvard.edu/abs/2003A&A...397..267A} {397, 267}

\bibitem[\protect\citeauthoryear{{Alecian} et~al.,}{{Alecian}
  et~al.}{2015}]{alecian:2015a}
{Alecian} E.,  et~al., 2015, in {Meynet} G.,  {Georgy} C.,  {Groh} J.,   {Stee}
  P.,  eds,  IAU Symposium Vol. 307, New Windows on Massive Stars. pp 330--335

\bibitem[\protect\citeauthoryear{{Alexeeva}, {Chen}, {Ryabchikova}, {Shi},
  {Sadakane}, {Nishimura}  \& {Zhao}}{{Alexeeva} et~al.}{2020}]{alexeeva:2020}
{Alexeeva} S.,  {Chen} T.,  {Ryabchikova} T.,  {Shi} W.,  {Sadakane} K.,
  {Nishimura} M.,   {Zhao} G.,  2020, \mn@doi [\apj]
  {10.3847/1538-4357/ab9306}, \href
  {https://ui.adsabs.harvard.edu/abs/2020ApJ...896...59A} {896, 59}

\bibitem[\protect\citeauthoryear{{Asplund}, {Grevesse}, {Sauval}  \&
  {Scott}}{{Asplund} et~al.}{2009}]{asplund:2009}
{Asplund} M.,  {Grevesse} N.,  {Sauval} A.~J.,   {Scott} P.,  2009, \mn@doi
  [\araa] {10.1146/annurev.astro.46.060407.145222}, \href
  {http://adsabs.harvard.edu/abs/2009ARA%26A..47..481A} {47, 481}

\bibitem[\protect\citeauthoryear{{Astropy Collaboration} et~al.,}{{Astropy
  Collaboration} et~al.}{2013}]{astropy:2013}
{Astropy Collaboration} et~al., 2013, \mn@doi [\aap]
  {10.1051/0004-6361/201322068}, \href
  {https://ui.adsabs.harvard.edu/abs/2013A&A...558A..33A} {558, A33}

\bibitem[\protect\citeauthoryear{{Astropy Collaboration} et~al.,}{{Astropy
  Collaboration} et~al.}{2018}]{astropy:2018}
{Astropy Collaboration} et~al., 2018, \mn@doi [\aj] {10.3847/1538-3881/aabc4f},
  \href {https://ui.adsabs.harvard.edu/abs/2018AJ....156..123A} {156, 123}

\bibitem[\protect\citeauthoryear{{Auri{\`e}re} et~al.,}{{Auri{\`e}re}
  et~al.}{2007}]{auriere:2007}
{Auri{\`e}re} M.,  et~al., 2007, \mn@doi [\aap] {10.1051/0004-6361:20078189},
  \href {http://adsabs.harvard.edu/abs/2007A%26A...475.1053A} {475, 1053}

\bibitem[\protect\citeauthoryear{{Auri{\`e}re} et~al.,}{{Auri{\`e}re}
  et~al.}{2010}]{auriere:2010a}
{Auri{\`e}re} M.,  et~al., 2010, \mn@doi [\aap] {10.1051/0004-6361/201014848},
  \href {http://adsabs.harvard.edu/abs/2010A%26A...523A..40A} {523, A40}

\bibitem[\protect\citeauthoryear{{Bagnulo}, {Landstreet}, {Mason}, {Andretta},
  {Silaj}  \& {Wade}}{{Bagnulo} et~al.}{2006}]{bagnulo:2006}
{Bagnulo} S.,  {Landstreet} J.~D.,  {Mason} E.,  {Andretta} V.,  {Silaj} J.,
  {Wade} G.~A.,  2006, \mn@doi [\aap] {10.1051/0004-6361:20054223}, \href
  {http://adsabs.harvard.edu/abs/2006A%26A...450..777B} {450, 777}

\bibitem[\protect\citeauthoryear{{Bagnulo}, {Landstreet}, {Fossati}  \&
  {Kochukhov}}{{Bagnulo} et~al.}{2012}]{bagnulo:2012}
{Bagnulo} S.,  {Landstreet} J.~D.,  {Fossati} L.,   {Kochukhov} O.,  2012,
  \mn@doi [\aap] {10.1051/0004-6361/201118098}, \href
  {http://adsabs.harvard.edu/abs/2012A%26A...538A.129B} {538, A129}

\bibitem[\protect\citeauthoryear{{Braithwaite} \& {Spruit}}{{Braithwaite} \&
  {Spruit}}{2004}]{braithwaite:2004}
{Braithwaite} J.,  {Spruit} H.~C.,  2004, \mn@doi [\nat] {10.1038/nature02934},
  \href {http://adsabs.harvard.edu/abs/2004Natur.431..819B} {431, 819}

\bibitem[\protect\citeauthoryear{{Brasseur}, {Phillip}, {Fleming}, {Mullally}
  \& {White}}{{Brasseur} et~al.}{2019}]{Brasseur2019}
{Brasseur} C.~E.,  {Phillip} C.,  {Fleming} S.~W.,  {Mullally} S.~E.,   {White}
  R.~L.,  2019, {Astrocut: Tools for creating cutouts of TESS images}
  (\mn@eprint {ascl} {1905.007})

\bibitem[\protect\citeauthoryear{{Briquet}, {Korhonen}, {Gonz{\'a}lez},
  {Hubrig}  \& {Hackman}}{{Briquet} et~al.}{2010}]{briquet:2010}
{Briquet} M.,  {Korhonen} H.,  {Gonz{\'a}lez} J.~F.,  {Hubrig} S.,   {Hackman}
  T.,  2010, \mn@doi [\aap] {10.1051/0004-6361/200913775}, \href
  {http://adsabs.harvard.edu/abs/2010A%26A...511A..71B} {511, A71}

\bibitem[\protect\citeauthoryear{{Carquillat} \& {Prieur}}{{Carquillat} \&
  {Prieur}}{2007}]{carquillat:2007}
{Carquillat} J.~M.,  {Prieur} J.~L.,  2007, \mn@doi [\mnras]
  {10.1111/j.1365-2966.2007.12143.x}, \href
  {https://ui.adsabs.harvard.edu/abs/2007MNRAS.380.1064C} {380, 1064}

\bibitem[\protect\citeauthoryear{{Choi}, {Dotter}, {Conroy}, {Cantiello},
  {Paxton}  \& {Johnson}}{{Choi} et~al.}{2016}]{choi:2016}
{Choi} J.,  {Dotter} A.,  {Conroy} C.,  {Cantiello} M.,  {Paxton} B.,
  {Johnson} B.~D.,  2016, \mn@doi [\apj] {10.3847/0004-637X/823/2/102}, \href
  {https://ui.adsabs.harvard.edu/abs/2016ApJ...823..102C} {823, 102}

\bibitem[\protect\citeauthoryear{{Claret}}{{Claret}}{2017}]{claret:2017}
{Claret} A.,  2017, \mn@doi [\aap] {10.1051/0004-6361/201629705}, \href
  {http://adsabs.harvard.edu/abs/2017A%26A...600A..30C} {600, A30}

\bibitem[\protect\citeauthoryear{{Claret} \& {Torres}}{{Claret} \&
  {Torres}}{2019}]{claret:2019}
{Claret} A.,  {Torres} G.,  2019, \mn@doi [\apj] {10.3847/1538-4357/ab1589},
  \href {https://ui.adsabs.harvard.edu/abs/2019ApJ...876..134C} {876, 134}

\bibitem[\protect\citeauthoryear{{Cowley}}{{Cowley}}{1972}]{cowley:1972}
{Cowley} A.,  1972, \mn@doi [\aj] {10.1086/111348}, \href
  {https://ui.adsabs.harvard.edu/abs/1972AJ.....77..750C} {77, 750}

\bibitem[\protect\citeauthoryear{{Cutri} et~al.,}{{Cutri}
  et~al.}{2003}]{cutri:2003}
{Cutri} R.~M.,  et~al., 2003, VizieR Online Data Catalog, \href
  {https://ui.adsabs.harvard.edu/abs/2003yCat.2246....0C} {p. II/246}

\bibitem[\protect\citeauthoryear{{Dotter}}{{Dotter}}{2016}]{dotter:2016}
{Dotter} A.,  2016, \mn@doi [\apjs] {10.3847/0067-0049/222/1/8}, \href
  {https://ui.adsabs.harvard.edu/abs/2016ApJS..222....8D} {222, 8}

\bibitem[\protect\citeauthoryear{{Dworetsky}}{{Dworetsky}}{1976}]{dworetsky:1976}
{Dworetsky} M.~M.,  1976, in {Weiss} W.~W.,  {Jenkner} H.,   {Wood} H.~J.,
  eds, IAU Colloq. 32: Physics of Ap Stars. p.~549

\bibitem[\protect\citeauthoryear{{Farbiash} \& {Steinitz}}{{Farbiash} \&
  {Steinitz}}{2004}]{farbiash:2004}
{Farbiash} N.,  {Steinitz} R.,  2004, in {Allen} C.,  {Scarfe} C.,  eds,
  Revista Mexicana de Astronomia y Astrofisica Conference Series Vol. 21,
  Revista Mexicana de Astronomia y Astrofisica Conference Series. pp 15--19

\bibitem[\protect\citeauthoryear{{Folsom}, {Kochukhov}, {Wade}, {Silvester}  \&
  {Bagnulo}}{{Folsom} et~al.}{2010}]{folsom:2010}
{Folsom} C.~P.,  {Kochukhov} O.,  {Wade} G.~A.,  {Silvester} J.,   {Bagnulo}
  S.,  2010, \mn@doi [\mnras] {10.1111/j.1365-2966.2010.17057.x}, \href
  {http://adsabs.harvard.edu/abs/2010MNRAS.407.2383F} {407, 2383}

\bibitem[\protect\citeauthoryear{Foreman-Mackey}{Foreman-Mackey}{2016}]{foreman-mackey:2016}
Foreman-Mackey D.,  2016, \mn@doi [The Journal of Open Source Software]
  {10.21105/joss.00024}, 1, 24

\bibitem[\protect\citeauthoryear{{Foreman-Mackey}, {Hogg}, {Lang}  \&
  {Goodman}}{{Foreman-Mackey} et~al.}{2013}]{foreman-mackey:2013}
{Foreman-Mackey} D.,  {Hogg} D.~W.,  {Lang} D.,   {Goodman} J.,  2013, \mn@doi
  [\pasp] {10.1086/670067}, \href
  {http://adsabs.harvard.edu/abs/2013PASP..125..306F} {125, 306}

\bibitem[\protect\citeauthoryear{{Foreman-Mackey} et~al.,}{{Foreman-Mackey}
  et~al.}{2019}]{foreman-mackey:2019}
{Foreman-Mackey} D.,  et~al., 2019, \mn@doi [The Journal of Open Source
  Software] {10.21105/joss.01864}, \href
  {https://ui.adsabs.harvard.edu/abs/2019JOSS....4.1864F} {4, 1864}

\bibitem[\protect\citeauthoryear{{Fossati}, {Ryabchikova}, {Bagnulo},
  {Alecian}, {Grunhut}, {Kochukhov}  \& {Wade}}{{Fossati}
  et~al.}{2009}]{fossati:2009}
{Fossati} L.,  {Ryabchikova} T.,  {Bagnulo} S.,  {Alecian} E.,  {Grunhut} J.,
  {Kochukhov} O.,   {Wade} G.,  2009, \mn@doi [\aap]
  {10.1051/0004-6361/200811561}, \href
  {http://cdsads.u-strasbg.fr/abs/2009A%26A...503..945F} {503, 945}

\bibitem[\protect\citeauthoryear{{Gaia Collaboration} et~al.,}{{Gaia
  Collaboration} et~al.}{2018}]{gaia-collaboration:2018}
{Gaia Collaboration} et~al., 2018, \mn@doi [\aap]
  {10.1051/0004-6361/201833051}, \href
  {https://ui.adsabs.harvard.edu/abs/2018A&A...616A...1G} {616, A1}

\bibitem[\protect\citeauthoryear{{Gandet}}{{Gandet}}{2008}]{gandet:2008}
{Gandet} T.~L.,  2008, Information Bulletin on Variable Stars, \href
  {https://ui.adsabs.harvard.edu/abs/2008IBVS.5848....1G} {5848, 1}

\bibitem[\protect\citeauthoryear{{Gerbaldi}, {Floquet}  \& {Hauck}}{{Gerbaldi}
  et~al.}{1985}]{gerbaldi:1985}
{Gerbaldi} M.,  {Floquet} M.,   {Hauck} B.,  1985, \aap, \href
  {http://adsabs.harvard.edu/abs/1985A%26A...146..341G} {146, 341}

\bibitem[\protect\citeauthoryear{{Ghazaryan}, {Alecian}  \&
  {Hakobyan}}{{Ghazaryan} et~al.}{2018}]{ghazaryan:2018}
{Ghazaryan} S.,  {Alecian} G.,   {Hakobyan} A.~A.,  2018, \mn@doi [\mnras]
  {10.1093/mnras/sty1912}, \href
  {https://ui.adsabs.harvard.edu/abs/2018MNRAS.480.2953G} {480, 2953}

\bibitem[\protect\citeauthoryear{{Gonz{\'a}lez}, {Hubrig}  \&
  {Castelli}}{{Gonz{\'a}lez} et~al.}{2010}]{gonzalez:2010a}
{Gonz{\'a}lez} J.~F.,  {Hubrig} S.,   {Castelli} F.,  2010, \mn@doi [\mnras]
  {10.1111/j.1365-2966.2009.16061.x}, \href
  {https://ui.adsabs.harvard.edu/abs/2010MNRAS.402.2539G} {402, 2539}

\bibitem[\protect\citeauthoryear{{Hale}}{{Hale}}{1994}]{hale:1994}
{Hale} A.,  1994, \mn@doi [\aj] {10.1086/116855}, \href
  {https://ui.adsabs.harvard.edu/abs/1994AJ....107..306H} {107, 306}

\bibitem[\protect\citeauthoryear{{Hauck} \& {North}}{{Hauck} \&
  {North}}{1982}]{hauck:1982}
{Hauck} B.,  {North} P.,  1982, \aap, \href
  {http://cdsads.u-strasbg.fr/abs/1982A%26A...114...23H} {114, 23}

\bibitem[\protect\citeauthoryear{{Hensberge} et~al.,}{{Hensberge}
  et~al.}{2007}]{hensberge:2007}
{Hensberge} H.,  et~al., 2007, \mn@doi [\mnras]
  {10.1111/j.1365-2966.2007.11955.x}, \href
  {https://ui.adsabs.harvard.edu/abs/2007MNRAS.379..349H} {379, 349}

\bibitem[\protect\citeauthoryear{{Huang}, {Gies}  \& {McSwain}}{{Huang}
  et~al.}{2010}]{huang:2010}
{Huang} W.,  {Gies} D.~R.,   {McSwain} M.~V.,  2010, \mn@doi [\apj]
  {10.1088/0004-637X/722/1/605}, \href
  {https://ui.adsabs.harvard.edu/abs/2010ApJ...722..605H} {722, 605}

\bibitem[\protect\citeauthoryear{{Hube}}{{Hube}}{1970}]{hube:1970}
{Hube} D.~P.,  1970, \memras, \href
  {https://ui.adsabs.harvard.edu/abs/1970MmRAS..72..233H} {72, 233}

\bibitem[\protect\citeauthoryear{{Hubrig} et~al.,}{{Hubrig}
  et~al.}{2012}]{hubrig:2012}
{Hubrig} S.,  et~al., 2012, \mn@doi [\aap] {10.1051/0004-6361/201219778}, \href
  {http://adsabs.harvard.edu/abs/2012A%26A...547A..90H} {547, A90}

\bibitem[\protect\citeauthoryear{{H{\"u}mmerich}, {Niemczura}, {Walczak},
  {Paunzen}, {Bernhard}, {Murphy}  \& {Drobek}}{{H{\"u}mmerich}
  et~al.}{2018}]{hummerich:2018}
{H{\"u}mmerich} S.,  {Niemczura} E.,  {Walczak} P.,  {Paunzen} E.,  {Bernhard}
  K.,  {Murphy} S.~J.,   {Drobek} D.,  2018, \mn@doi [\mnras]
  {10.1093/mnras/stx2974}, \href
  {https://ui.adsabs.harvard.edu/abs/2018MNRAS.474.2467H} {474, 2467}

\bibitem[\protect\citeauthoryear{{Hut}}{{Hut}}{1981}]{hut:1981}
{Hut} P.,  1981, \aap, \href
  {https://ui.adsabs.harvard.edu/abs/1981A&A....99..126H} {99, 126}

\bibitem[\protect\citeauthoryear{{Johnston}, {Tkachenko}, {Aerts},
  {Molenberghs}, {Bowman}, {Pedersen}, {Buysschaert}  \&
  {P{\'a}pics}}{{Johnston} et~al.}{2019}]{johnston:2019}
{Johnston} C.,  {Tkachenko} A.,  {Aerts} C.,  {Molenberghs} G.,  {Bowman}
  D.~M.,  {Pedersen} M.~G.,  {Buysschaert} B.,   {P{\'a}pics} P.~I.,  2019,
  \mn@doi [\mnras] {10.1093/mnras/sty2671}, \href
  {https://ui.adsabs.harvard.edu/abs/2019MNRAS.482.1231J} {482, 1231}

\bibitem[\protect\citeauthoryear{{Kazarovets}, {Samus}, {Durlevich}, {Frolov},
  {Antipin}, {Kireeva}  \& {Pastukhova}}{{Kazarovets}
  et~al.}{1999}]{kazarovets:1999}
{Kazarovets} E.~V.,  {Samus} N.~N.,  {Durlevich} O.~V.,  {Frolov} M.~S.,
  {Antipin} S.~V.,  {Kireeva} N.~N.,   {Pastukhova} E.~N.,  1999, Information
  Bulletin on Variable Stars, \href
  {https://ui.adsabs.harvard.edu/abs/1999IBVS.4659....1K} {4659, 1}

\bibitem[\protect\citeauthoryear{{Kochukhov}}{{Kochukhov}}{2005}]{kochukhov:2005}
{Kochukhov} O.,  2005, \mn@doi [\aap] {10.1051/0004-6361:20052629}, \href
  {http://adsabs.harvard.edu/abs/2005A%26A...438..219K} {438, 219}

\bibitem[\protect\citeauthoryear{{Kochukhov}}{{Kochukhov}}{2007}]{kochukhov:2007d}
{Kochukhov} O.,  2007, in {Romanyuk} I.~I.,  {Kudryavtsev} D.~O.,  eds, Physics
  of Magnetic Stars. pp 109--118

\bibitem[\protect\citeauthoryear{{Kochukhov}}{{Kochukhov}}{2018}]{kochukhov:2018}
{Kochukhov} O.,  2018, {BinMag: Widget for comparing stellar observed with
  theoretical spectra}, Astrophysics Source Code Library (\mn@eprint {ascl}
  {1805.015})

\bibitem[\protect\citeauthoryear{{Kochukhov}, {Adelman}, {Gulliver}  \&
  {Piskunov}}{{Kochukhov} et~al.}{2007}]{kochukhov:2007b}
{Kochukhov} O.,  {Adelman} S.~J.,  {Gulliver} A.~F.,   {Piskunov} N.,  2007,
  \mn@doi [Nature Physics] {10.1038/nphys648}, \href
  {http://esoads.eso.org/abs/2007NatPh...3..526K} {3, 526}

\bibitem[\protect\citeauthoryear{{Kochukhov}, {Makaganiuk}  \&
  {Piskunov}}{{Kochukhov} et~al.}{2010}]{kochukhov:2010a}
{Kochukhov} O.,  {Makaganiuk} V.,   {Piskunov} N.,  2010, \mn@doi [\aap]
  {10.1051/0004-6361/201015429}, \href
  {https://ui.adsabs.harvard.edu/abs/2010A&A...524A...5K} {524, A5}

\bibitem[\protect\citeauthoryear{{Kochukhov} et~al.,}{{Kochukhov}
  et~al.}{2011}]{kochukhov:2011b}
{Kochukhov} O.,  et~al., 2011, \mn@doi [\aap] {10.1051/0004-6361/201117970},
  \href {http://cdsads.u-strasbg.fr/abs/2011A%26A...534L..13K} {534, L13}

\bibitem[\protect\citeauthoryear{{Kochukhov} et~al.,}{{Kochukhov}
  et~al.}{2013}]{kochukhov:2013a}
{Kochukhov} O.,  et~al., 2013, \mn@doi [\aap] {10.1051/0004-6361/201321467},
  \href {http://adsabs.harvard.edu/abs/2013A%26A...554A..61K} {554, A61}

\bibitem[\protect\citeauthoryear{{Kochukhov}, {Johnston}, {Alecian}  \&
  {Wade}}{{Kochukhov} et~al.}{2018}]{kochukhov:2018b}
{Kochukhov} O.,  {Johnston} C.,  {Alecian} E.,   {Wade} G.~A.,  2018, \mn@doi
  [\mnras] {10.1093/mnras/sty1118}, \href
  {http://adsabs.harvard.edu/abs/2018MNRAS.478.1749K} {478, 1749}

\bibitem[\protect\citeauthoryear{{Kochukhov}, {Shultz}  \&
  {Neiner}}{{Kochukhov} et~al.}{2019}]{kochukhov:2019}
{Kochukhov} O.,  {Shultz} M.,   {Neiner} C.,  2019, \mn@doi [\aap]
  {10.1051/0004-6361/201834279}, \href
  {http://adsabs.harvard.edu/abs/2019A%26A...621A..47K} {621, A47}

\bibitem[\protect\citeauthoryear{{Korhonen} et~al.,}{{Korhonen}
  et~al.}{2013}]{korhonen:2013}
{Korhonen} H.,  et~al., 2013, \mn@doi [\aap] {10.1051/0004-6361/201220951},
  \href {http://adsabs.harvard.edu/abs/2013A%26A...553A..27K} {553, A27}

\bibitem[\protect\citeauthoryear{{Kunzli}, {North}, {Kurucz}  \&
  {Nicolet}}{{Kunzli} et~al.}{1997}]{kunzli:1997}
{Kunzli} M.,  {North} P.,  {Kurucz} R.~L.,   {Nicolet} B.,  1997, \aaps, \href
  {http://adsabs.harvard.edu/abs/1997A%26AS..122...51K} {122, 51}

\bibitem[\protect\citeauthoryear{{Lallement}, {Babusiaux}, {Vergely}, {Katz},
  {Arenou}, {Valette}, {Hottier}  \& {Capitanio}}{{Lallement}
  et~al.}{2019}]{lallement:2019}
{Lallement} R.,  {Babusiaux} C.,  {Vergely} J.~L.,  {Katz} D.,  {Arenou} F.,
  {Valette} B.,  {Hottier} C.,   {Capitanio} L.,  2019, \mn@doi [\aap]
  {10.1051/0004-6361/201834695}, \href
  {https://ui.adsabs.harvard.edu/abs/2019A&A...625A.135L} {625, A135}

\bibitem[\protect\citeauthoryear{{Landstreet}, {Bagnulo}, {Andretta},
  {Fossati}, {Mason}, {Silaj}  \& {Wade}}{{Landstreet}
  et~al.}{2007}]{landstreet:2007}
{Landstreet} J.~D.,  {Bagnulo} S.,  {Andretta} V.,  {Fossati} L.,  {Mason} E.,
  {Silaj} J.,   {Wade} G.~A.,  2007, \mn@doi [\aap]
  {10.1051/0004-6361:20077343}, \href
  {http://adsabs.harvard.edu/abs/2007A%26A...470..685L} {470, 685}

\bibitem[\protect\citeauthoryear{{Landstreet}, {Kochukhov}, {Alecian},
  {Bailey}, {Mathis}, {Neiner}, {Wade}  \& {BINaMIcS
  Collaboration}}{{Landstreet} et~al.}{2017}]{landstreet:2017}
{Landstreet} J.~D.,  {Kochukhov} O.,  {Alecian} E.,  {Bailey} J.~D.,  {Mathis}
  S.,  {Neiner} C.,  {Wade} G.~A.,   {BINaMIcS Collaboration} 2017, \mn@doi
  [\aap] {10.1051/0004-6361/201630233}, \href
  {http://adsabs.harvard.edu/abs/2017A%26A...601A.129L} {601, A129}

\bibitem[\protect\citeauthoryear{{Lightkurve Collaboration}
  et~al.,}{{Lightkurve Collaboration} et~al.}{2018}]{Lightkurve2018}
{Lightkurve Collaboration} et~al., 2018, {Lightkurve: Kepler and TESS time
  series analysis in Python}, Astrophysics Source Code Library (\mn@eprint
  {ascl} {1812.013})

\bibitem[\protect\citeauthoryear{{Makaganiuk} et~al.,}{{Makaganiuk}
  et~al.}{2011a}]{makaganiuk:2011a}
{Makaganiuk} V.,  et~al., 2011a, \mn@doi [\aap] {10.1051/0004-6361/201015666},
  \href {http://adsabs.harvard.edu/abs/2011A%26A...525A..97M} {525, A97}

\bibitem[\protect\citeauthoryear{{Makaganiuk} et~al.,}{{Makaganiuk}
  et~al.}{2011b}]{makaganiuk:2011}
{Makaganiuk} V.,  et~al., 2011b, \mn@doi [\aap] {10.1051/0004-6361/201016302},
  \href {http://adsabs.harvard.edu/abs/2011A%26A...529A.160M} {529, A160}

\bibitem[\protect\citeauthoryear{{Makaganiuk} et~al.,}{{Makaganiuk}
  et~al.}{2012}]{makaganiuk:2012}
{Makaganiuk} V.,  et~al., 2012, \mn@doi [\aap] {10.1051/0004-6361/201118167},
  \href {http://adsabs.harvard.edu/abs/2012A%26A...539A.142M} {539, A142}

\bibitem[\protect\citeauthoryear{{Mashonkina}}{{Mashonkina}}{2020}]{mashonkina:2020}
{Mashonkina} L.,  2020, \mn@doi [\mnras] {10.1093/mnras/staa653}, \href
  {https://ui.adsabs.harvard.edu/abs/2020MNRAS.493.6095M} {493, 6095}

\bibitem[\protect\citeauthoryear{{Mathys}}{{Mathys}}{2017}]{mathys:2017}
{Mathys} G.,  2017, \mn@doi [\aap] {10.1051/0004-6361/201628429}, \href
  {http://adsabs.harvard.edu/abs/2017A%26A...601A..14M} {601, A14}

\bibitem[\protect\citeauthoryear{{Maxted} et~al.,}{{Maxted}
  et~al.}{2020}]{maxted:2020}
{Maxted} P.~F.~L.,  et~al., 2020, \mn@doi [\mnras] {10.1093/mnras/staa1662},
  \href {https://ui.adsabs.harvard.edu/abs/2020MNRAS.tmp.1795M} {}

\bibitem[\protect\citeauthoryear{{Mestel}}{{Mestel}}{1999}]{mestel:1999}
{Mestel} L.,  1999, {Stellar magnetism}.
Oxford Science Publications

\bibitem[\protect\citeauthoryear{{Michaud}}{{Michaud}}{1982}]{michaud:1982}
{Michaud} G.,  1982, \mn@doi [\apj] {10.1086/160083}, \href
  {https://ui.adsabs.harvard.edu/abs/1982ApJ...258..349M} {258, 349}

\bibitem[\protect\citeauthoryear{{Michaud}, {Alecian}  \& {Richer}}{{Michaud}
  et~al.}{2015}]{michaud:2015}
{Michaud} G.,  {Alecian} G.,   {Richer} J.,  2015, {Atomic Diffusion in Stars},
  \mn@doi{10.1007/978-3-319-19854-5.
}

\bibitem[\protect\citeauthoryear{{Morel} et~al.,}{{Morel}
  et~al.}{2014}]{morel:2014}
{Morel} T.,  et~al., 2014, \mn@doi [\aap] {10.1051/0004-6361/201322289}, \href
  {http://adsabs.harvard.edu/abs/2014A%26A...561A..35M} {561, A35}

\bibitem[\protect\citeauthoryear{{Moss}}{{Moss}}{2004}]{moss:2004}
{Moss} D.,  2004, in {Zverko} J.,  {Ziznovsky} J.,  {Adelman} S.~J.,   {Weiss}
  W.~W.,  eds,  IAU Symposium Vol. 224, The A-Star Puzzle. pp 245--252

\bibitem[\protect\citeauthoryear{{Neiner}, {Mathis}, {Alecian}, {Emeriau},
  {Grunhut}, {BinaMIcS}  \& {MiMeS Collaborations}}{{Neiner}
  et~al.}{2015}]{neiner:2015}
{Neiner} C.,  {Mathis} S.,  {Alecian} E.,  {Emeriau} C.,  {Grunhut} J.,
  {BinaMIcS}  {MiMeS Collaborations} 2015, in {Nagendra} K.~N.,  {Bagnulo} S.,
  {Centeno} R.,   {Jes{\'u}s Mart{\'{\i}}nez Gonz{\'a}lez} M.,  eds,  IAU
  Symposium Vol. 305, Polarimetry. pp 61--66

\bibitem[\protect\citeauthoryear{{Nordstrom} \& {Johansen}}{{Nordstrom} \&
  {Johansen}}{1994}]{nordstrom:1994}
{Nordstrom} B.,  {Johansen} K.~T.,  1994, \aap, \href
  {http://adsabs.harvard.edu/abs/1994A%26A...282..787N} {282, 787}

\bibitem[\protect\citeauthoryear{{Otero}}{{Otero}}{2007}]{otero:2007}
{Otero} S.~A.,  2007, Open European Journal on Variable Stars, \href
  {https://ui.adsabs.harvard.edu/abs/2007OEJV...72....1O} {0072, 1}

\bibitem[\protect\citeauthoryear{{Pakhomov}, {Ryabchikova}  \&
  {Piskunov}}{{Pakhomov} et~al.}{2019}]{pakhomov:2019}
{Pakhomov} Y.~V.,  {Ryabchikova} T.~A.,   {Piskunov} N.~E.,  2019, \mn@doi
  [Astronomy Reports] {10.1134/S1063772919120047}, \href
  {https://ui.adsabs.harvard.edu/abs/2019ARep...63.1010P} {63, 1010}

\bibitem[\protect\citeauthoryear{{Paunzen}, {Schnell}  \& {Maitzen}}{{Paunzen}
  et~al.}{2005}]{paunzen:2005a}
{Paunzen} E.,  {Schnell} A.,   {Maitzen} H.~M.,  2005, \mn@doi [\aap]
  {10.1051/0004-6361:20053546}, \href
  {https://ui.adsabs.harvard.edu/abs/2005A&A...444..941P} {444, 941}

\bibitem[\protect\citeauthoryear{{Paunzen}, {Schnell}  \& {Maitzen}}{{Paunzen}
  et~al.}{2006}]{paunzen:2006}
{Paunzen} E.,  {Schnell} A.,   {Maitzen} H.~M.,  2006, \mn@doi [\aap]
  {10.1051/0004-6361:20064889}, \href
  {https://ui.adsabs.harvard.edu/abs/2006A&A...458..293P} {458, 293}

\bibitem[\protect\citeauthoryear{{Perryman} et~al.,}{{Perryman}
  et~al.}{1997}]{perryman:1997}
{Perryman} M.~A.~C.,  et~al., 1997, \aap, \href
  {http://adsabs.harvard.edu/abs/1997A%26A...323L..49P} {323, L49}

\bibitem[\protect\citeauthoryear{{Prsa}, {Matijevic}, {Latkovic}, {Vilardell}
  \& {Wils}}{{Prsa} et~al.}{2011}]{prsa:2011}
{Prsa} A.,  {Matijevic} G.,  {Latkovic} O.,  {Vilardell} F.,   {Wils} P.,
  2011, {PHOEBE: PHysics Of Eclipsing BinariEs}, Astrophysics Source Code
  Library (\mn@eprint {ascl} {1106.002})

\bibitem[\protect\citeauthoryear{{Pr{\v s}a} \& {Zwitter}}{{Pr{\v s}a} \&
  {Zwitter}}{2005}]{prsa:2005}
{Pr{\v s}a} A.,  {Zwitter} T.,  2005, \mn@doi [\apj] {10.1086/430591}, \href
  {http://adsabs.harvard.edu/abs/2005ApJ...628..426P} {628, 426}

\bibitem[\protect\citeauthoryear{{Pr{\v s}a} et~al.,}{{Pr{\v s}a}
  et~al.}{2016}]{prsa:2016}
{Pr{\v s}a} A.,  et~al., 2016, \mn@doi [\apjs] {10.3847/1538-4365/227/2/29},
  \href {http://adsabs.harvard.edu/abs/2016ApJS..227...29P} {227, 29}

\bibitem[\protect\citeauthoryear{{Prv{\'a}k}, {Krti{\v{c}}ka}  \&
  {Korhonen}}{{Prv{\'a}k} et~al.}{2020}]{prvak:2020}
{Prv{\'a}k} M.,  {Krti{\v{c}}ka} J.,   {Korhonen} H.,  2020, \mn@doi [\mnras]
  {10.1093/mnras/stz3564}, \href
  {https://ui.adsabs.harvard.edu/abs/2020MNRAS.492.1834P} {492, 1834}

\bibitem[\protect\citeauthoryear{{Raskin} et~al.,}{{Raskin}
  et~al.}{2011}]{raskin:2011}
{Raskin} G.,  et~al., 2011, \mn@doi [\aap] {10.1051/0004-6361/201015435}, \href
  {https://ui.adsabs.harvard.edu/abs/2011A&A...526A..69R} {526, A69}

\bibitem[\protect\citeauthoryear{{Renson} \& {Manfroid}}{{Renson} \&
  {Manfroid}}{2009}]{renson:2009}
{Renson} P.,  {Manfroid} J.,  2009, \mn@doi [\aap]
  {10.1051/0004-6361/200810788}, \href
  {http://adsabs.harvard.edu/abs/2009A%26A...498..961R} {498, 961}

\bibitem[\protect\citeauthoryear{{Richard}, {Michaud}  \& {Richer}}{{Richard}
  et~al.}{2001}]{richard:2001}
{Richard} O.,  {Michaud} G.,   {Richer} J.,  2001, \mn@doi [\apj]
  {10.1086/322264}, \href
  {https://ui.adsabs.harvard.edu/abs/2001ApJ...558..377R} {558, 377}

\bibitem[\protect\citeauthoryear{{Ricker} et~al.,}{{Ricker}
  et~al.}{2015}]{ricker:2015}
{Ricker} G.~R.,  et~al., 2015, \mn@doi [Journal of Astronomical Telescopes,
  Instruments, and Systems] {10.1117/1.JATIS.1.1.014003}, \href
  {http://adsabs.harvard.edu/abs/2015JATIS...1a4003R} {1, 014003}

\bibitem[\protect\citeauthoryear{{Ros{\'e}n}, {Kochukhov}, {Alecian}, {Neiner},
  {Morin}, {Wade}  \& {BinaMIcS Collaboration}}{{Ros{\'e}n}
  et~al.}{2018}]{rosen:2018}
{Ros{\'e}n} L.,  {Kochukhov} O.,  {Alecian} E.,  {Neiner} C.,  {Morin} J.,
  {Wade} G.~A.,   {BinaMIcS Collaboration} 2018, \mn@doi [\aap]
  {10.1051/0004-6361/201731706}, \href
  {http://adsabs.harvard.edu/abs/2018A%26A...613A..60R} {613, A60}

\bibitem[\protect\citeauthoryear{{Rusomarov}, {Kochukhov}, {Ryabchikova}  \&
  {Ilyin}}{{Rusomarov} et~al.}{2016}]{rusomarov:2016}
{Rusomarov} N.,  {Kochukhov} O.,  {Ryabchikova} T.,   {Ilyin} I.,  2016,
  \mn@doi [\aap] {10.1051/0004-6361/201527719}, \href
  {http://adsabs.harvard.edu/abs/2016A%26A...588A.138R} {588, A138}

\bibitem[\protect\citeauthoryear{{Ryabchikova}}{{Ryabchikova}}{1998}]{ryabchikova:1998a}
{Ryabchikova} T.,  1998, Contributions of the Astronomical Observatory Skalnate
  Pleso, \href {https://ui.adsabs.harvard.edu/abs/1998CoSka..27..319R} {27,
  319}

\bibitem[\protect\citeauthoryear{{Ryabchikova}, {Malanushenko}  \&
  {Adelman}}{{Ryabchikova} et~al.}{1999}]{ryabchikova:1999a}
{Ryabchikova} T.~A.,  {Malanushenko} V.~P.,   {Adelman} S.~J.,  1999, \aap,
  \href {http://adsabs.harvard.edu/abs/1999A%26A...351..963R} {351, 963}

\bibitem[\protect\citeauthoryear{{Ryabchikova}, {Piskunov}, {Kurucz},
  {Stempels}, {Heiter}, {Pakhomov}  \& {Barklem}}{{Ryabchikova}
  et~al.}{2015}]{ryabchikova:2015}
{Ryabchikova} T.,  {Piskunov} N.,  {Kurucz} R.~L.,  {Stempels} H.~C.,  {Heiter}
  U.,  {Pakhomov} Y.,   {Barklem} P.~S.,  2015, \mn@doi [\physscr]
  {10.1088/0031-8949/90/5/054005}, \href
  {http://adsabs.harvard.edu/abs/2015PhyS...90e4005R} {90, 054005}

\bibitem[\protect\citeauthoryear{{Samus'}, {Kazarovets}, {Durlevich}, {Kireeva}
   \& {Pastukhova}}{{Samus'} et~al.}{2017}]{samus:2017}
{Samus'} N.~N.,  {Kazarovets} E.~V.,  {Durlevich} O.~V.,  {Kireeva} N.~N.,
  {Pastukhova} E.~N.,  2017, \mn@doi [Astronomy Reports]
  {10.1134/S1063772917010085}, \href
  {https://ui.adsabs.harvard.edu/abs/2017ARep...61...80S} {61, 80}

\bibitem[\protect\citeauthoryear{{Schneider}, {Podsiadlowski}, {Langer},
  {Castro}  \& {Fossati}}{{Schneider} et~al.}{2016}]{schneider:2016}
{Schneider} F.~R.~N.,  {Podsiadlowski} P.,  {Langer} N.,  {Castro} N.,
  {Fossati} L.,  2016, \mn@doi [\mnras] {10.1093/mnras/stw148}, \href
  {http://adsabs.harvard.edu/abs/2016MNRAS.457.2355S} {457, 2355}

\bibitem[\protect\citeauthoryear{{Schneider}, {Ohlmann}, {Podsiadlowski},
  {R{\"o}pke}, {Balbus}, {Pakmor}  \& {Springel}}{{Schneider}
  et~al.}{2019}]{schneider:2019}
{Schneider} F. R.~N.,  {Ohlmann} S.~T.,  {Podsiadlowski} P.,  {R{\"o}pke}
  F.~K.,  {Balbus} S.~A.,  {Pakmor} R.,   {Springel} V.,  2019, \mn@doi [\nat]
  {10.1038/s41586-019-1621-5}, \href
  {https://ui.adsabs.harvard.edu/abs/2019Natur.574..211S} {574, 211}

\bibitem[\protect\citeauthoryear{{Shorlin}, {Wade}, {Donati}, {Landstreet},
  {Petit}, {Sigut}  \& {Strasser}}{{Shorlin} et~al.}{2002}]{shorlin:2002}
{Shorlin} S.~L.~S.,  {Wade} G.~A.,  {Donati} J.-F.,  {Landstreet} J.~D.,
  {Petit} P.,  {Sigut} T.~A.~A.,   {Strasser} S.,  2002, \mn@doi [\aap]
  {10.1051/0004-6361:20021192}, \href
  {http://adsabs.harvard.edu/abs/2002A%26A...392..637S} {392, 637}

\bibitem[\protect\citeauthoryear{{Shultz} et~al.,}{{Shultz}
  et~al.}{2019}]{shultz:2019}
{Shultz} M.~E.,  et~al., 2019, \mn@doi [\mnras] {10.1093/mnras/stz2846}, \href
  {https://ui.adsabs.harvard.edu/abs/2019MNRAS.490.4154S} {490, 4154}

\bibitem[\protect\citeauthoryear{{Shulyak}, {Tsymbal}, {Ryabchikova},
  {St{\"u}tz}  \& {Weiss}}{{Shulyak} et~al.}{2004}]{shulyak:2004}
{Shulyak} D.,  {Tsymbal} V.,  {Ryabchikova} T.,  {St{\"u}tz} C.,   {Weiss}
  W.~W.,  2004, \mn@doi [\aap] {10.1051/0004-6361:20034169}, \href
  {http://adsabs.harvard.edu/abs/2004A%26A...428..993S} {428, 993}

\bibitem[\protect\citeauthoryear{{Sikora}, {Wade}, {Power}  \&
  {Neiner}}{{Sikora} et~al.}{2019}]{sikora:2019a}
{Sikora} J.,  {Wade} G.~A.,  {Power} J.,   {Neiner} C.,  2019, \mn@doi [\mnras]
  {10.1093/mnras/sty2895}, \href
  {https://ui.adsabs.harvard.edu/abs/2019MNRAS.483.3127S} {483, 3127}

\bibitem[\protect\citeauthoryear{{Sitnova}, {Mashonkina}  \&
  {Ryabchikova}}{{Sitnova} et~al.}{2018}]{sitnova:2018}
{Sitnova} T.~M.,  {Mashonkina} L.~I.,   {Ryabchikova} T.~A.,  2018, \mn@doi
  [\mnras] {10.1093/mnras/sty810}, \href
  {https://ui.adsabs.harvard.edu/abs/2018MNRAS.477.3343S} {477, 3343}

\bibitem[\protect\citeauthoryear{{Skarka} et~al.,}{{Skarka}
  et~al.}{2019}]{skarka:2019}
{Skarka} M.,  et~al., 2019, \mn@doi [\mnras] {10.1093/mnras/stz1478}, \href
  {https://ui.adsabs.harvard.edu/abs/2019MNRAS.487.4230S} {487, 4230}

\bibitem[\protect\citeauthoryear{{Sofia} \& {Meyer}}{{Sofia} \&
  {Meyer}}{2001}]{sofia:2001}
{Sofia} U.~J.,  {Meyer} D.~M.,  2001, \mn@doi [\apjl] {10.1086/321715}, \href
  {https://ui.adsabs.harvard.edu/abs/2001ApJ...554L.221S} {554, L221}

\bibitem[\protect\citeauthoryear{{Strassmeier}, {Granzer}, {Mallonn}, {Weber}
  \& {Weingrill}}{{Strassmeier} et~al.}{2017}]{strassmeier:2017}
{Strassmeier} K.~G.,  {Granzer} T.,  {Mallonn} M.,  {Weber} M.,   {Weingrill}
  J.,  2017, \mn@doi [\aap] {10.1051/0004-6361/201629150}, \href
  {http://adsabs.harvard.edu/abs/2017A%26A...597A..55S} {597, A55}

\bibitem[\protect\citeauthoryear{{Takeda}, {Han}, {Kang}, {Lee}  \&
  {Kim}}{{Takeda} et~al.}{2019}]{takeda:2019}
{Takeda} Y.,  {Han} I.,  {Kang} D.-I.,  {Lee} B.-C.,   {Kim} K.-M.,  2019,
  \mn@doi [\mnras] {10.1093/mnras/stz449}, \href
  {https://ui.adsabs.harvard.edu/abs/2019MNRAS.485.1067T} {485, 1067}

\bibitem[\protect\citeauthoryear{{Thompson}, {Nandy}, {Jamar}, {Monfils},
  {Houziaux}, {Carnochan}  \& {Wilson}}{{Thompson}
  et~al.}{1978}]{thompson:1978}
{Thompson} G.~I.,  {Nandy} K.,  {Jamar} C.,  {Monfils} A.,  {Houziaux} L.,
  {Carnochan} D.~J.,   {Wilson} R.,  1978, {Catalogue of stellar ultraviolet
  fluxes : a compilation of absolute stellar fluxes measured by the Sky Survey
  Telescope (S2/68) aboard the ESRO satellite TD-1}

\bibitem[\protect\citeauthoryear{{Torres}, {Andersen}  \&
  {Gim{\'e}nez}}{{Torres} et~al.}{2010}]{torres:2010}
{Torres} G.,  {Andersen} J.,   {Gim{\'e}nez} A.,  2010, \mn@doi [\aapr]
  {10.1007/s00159-009-0025-1}, \href
  {http://adsabs.harvard.edu/abs/2010A%26ARv..18...67T} {18, 67}

\bibitem[\protect\citeauthoryear{{Vidotto} et~al.,}{{Vidotto}
  et~al.}{2014}]{vidotto:2014}
{Vidotto} A.~A.,  et~al., 2014, \mn@doi [\mnras] {10.1093/mnras/stu728}, \href
  {https://ui.adsabs.harvard.edu/abs/2014MNRAS.441.2361V} {441, 2361}

\bibitem[\protect\citeauthoryear{{White} et~al.,}{{White}
  et~al.}{2017}]{white:2017}
{White} T.~R.,  et~al., 2017, \mn@doi [\mnras] {10.1093/mnras/stx1050}, \href
  {https://ui.adsabs.harvard.edu/abs/2017MNRAS.471.2882W} {471, 2882}

\bibitem[\protect\citeauthoryear{{Woolf} \& {Lambert}}{{Woolf} \&
  {Lambert}}{1999}]{woolf:1999}
{Woolf} V.~M.,  {Lambert} D.~L.,  1999, \mn@doi [\apj] {10.1086/307551}, \href
  {http://adsabs.harvard.edu/abs/1999ApJ...521..414W} {521, 414}

\bibitem[\protect\citeauthoryear{{Zahn}}{{Zahn}}{1977}]{zahn:1977}
{Zahn} J.-P.,  1977, \aap, \href
  {http://adsabs.harvard.edu/abs/1977A%26A....57..383Z} {57, 383}

\bibitem[\protect\citeauthoryear{{de Mink}, {Sana}, {Langer}, {Izzard}  \&
  {Schneider}}{{de Mink} et~al.}{2014}]{de-mink:2014}
{de Mink} S.~E.,  {Sana} H.,  {Langer} N.,  {Izzard} R.~G.,   {Schneider}
  F.~R.~N.,  2014, \mn@doi [\apj] {10.1088/0004-637X/782/1/7}, \href
  {https://ui.adsabs.harvard.edu/abs/2014ApJ...782....7D} {782, 7}

\makeatother
\end{thebibliography}

\newpage

\appendix

\section[]{Posterior probability distributions}

\begin{figure}
\centering
\figps{\hsize}{0}{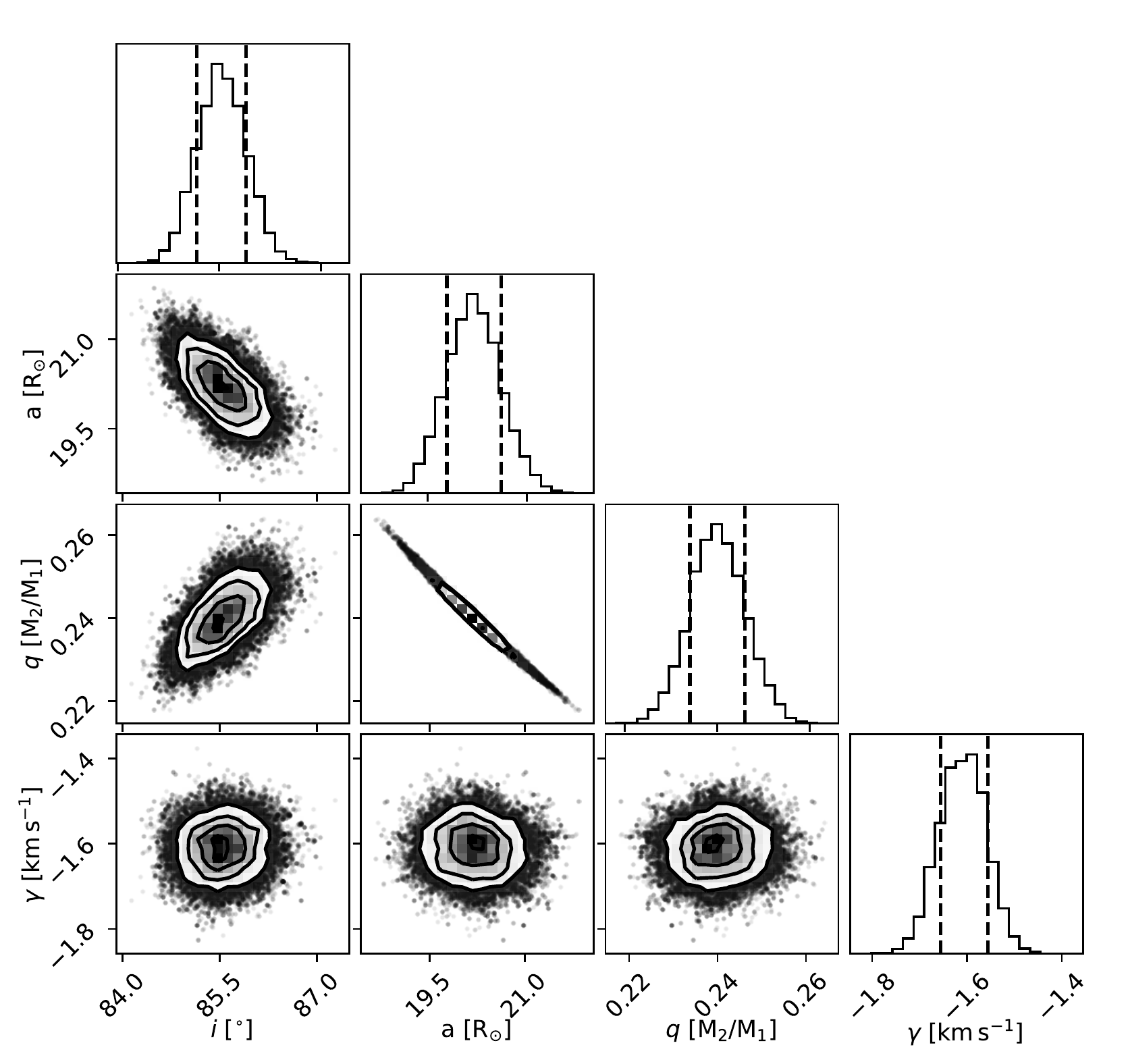}
\caption{Marginalised posterior distributions for orbital parameters. The dashed lines in histogram plots denote 68 per cent (1$\sigma$) confidence ranges. The contours in the probability density plots correspond to 0.5, 1.0, 1.5, and 2.0$\sigma$ confidence intervals.}
\label{fig:pd1}
\end{figure}

\begin{figure}
\centering
\figps{\hsize}{0}{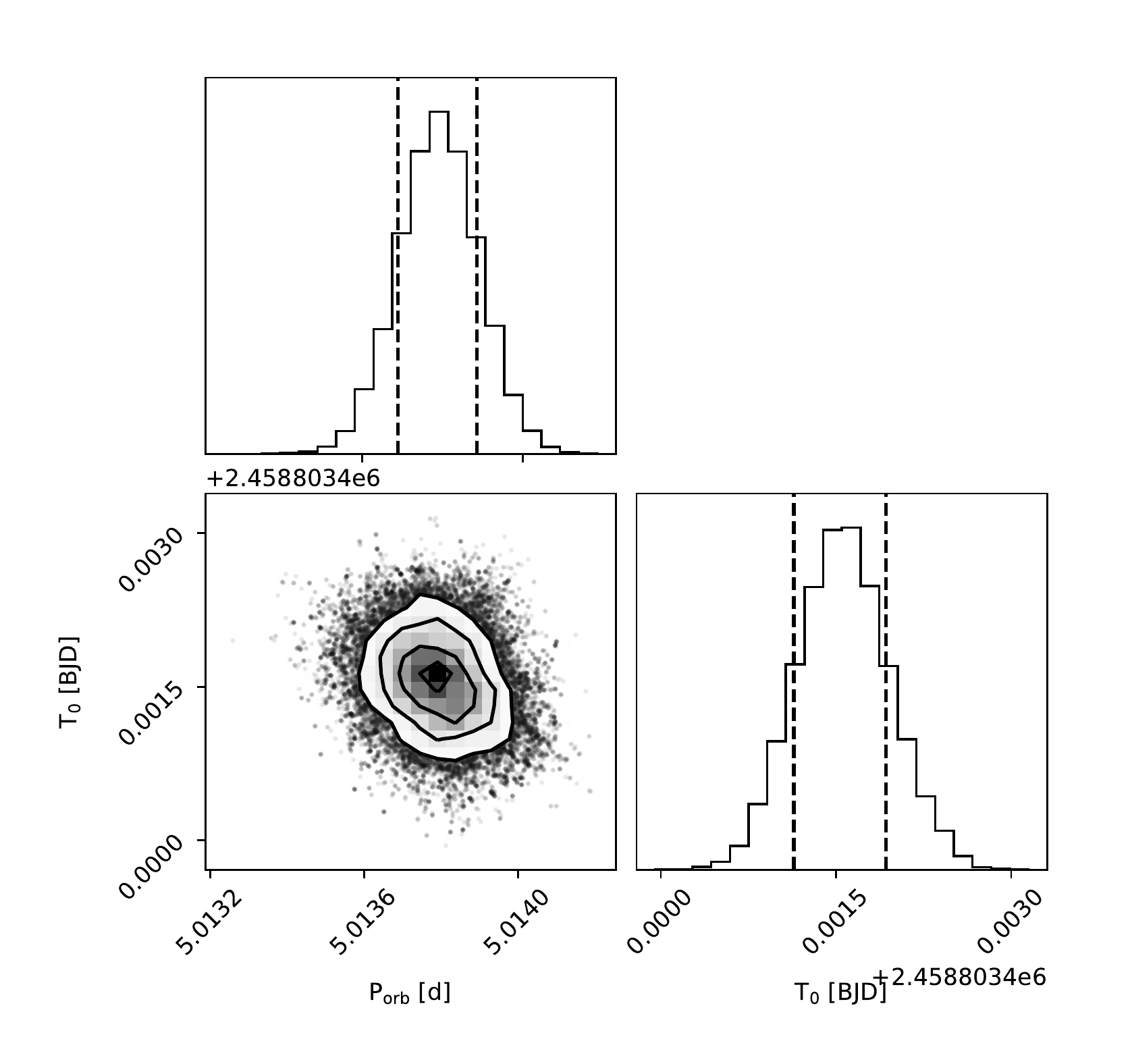}
\caption{Same as Fig.~\ref{fig:pd1}, but for the orbital ephemeris.}
\label{fig:pd2}
\end{figure}

\begin{figure}
\centering
\figps{\hsize}{0}{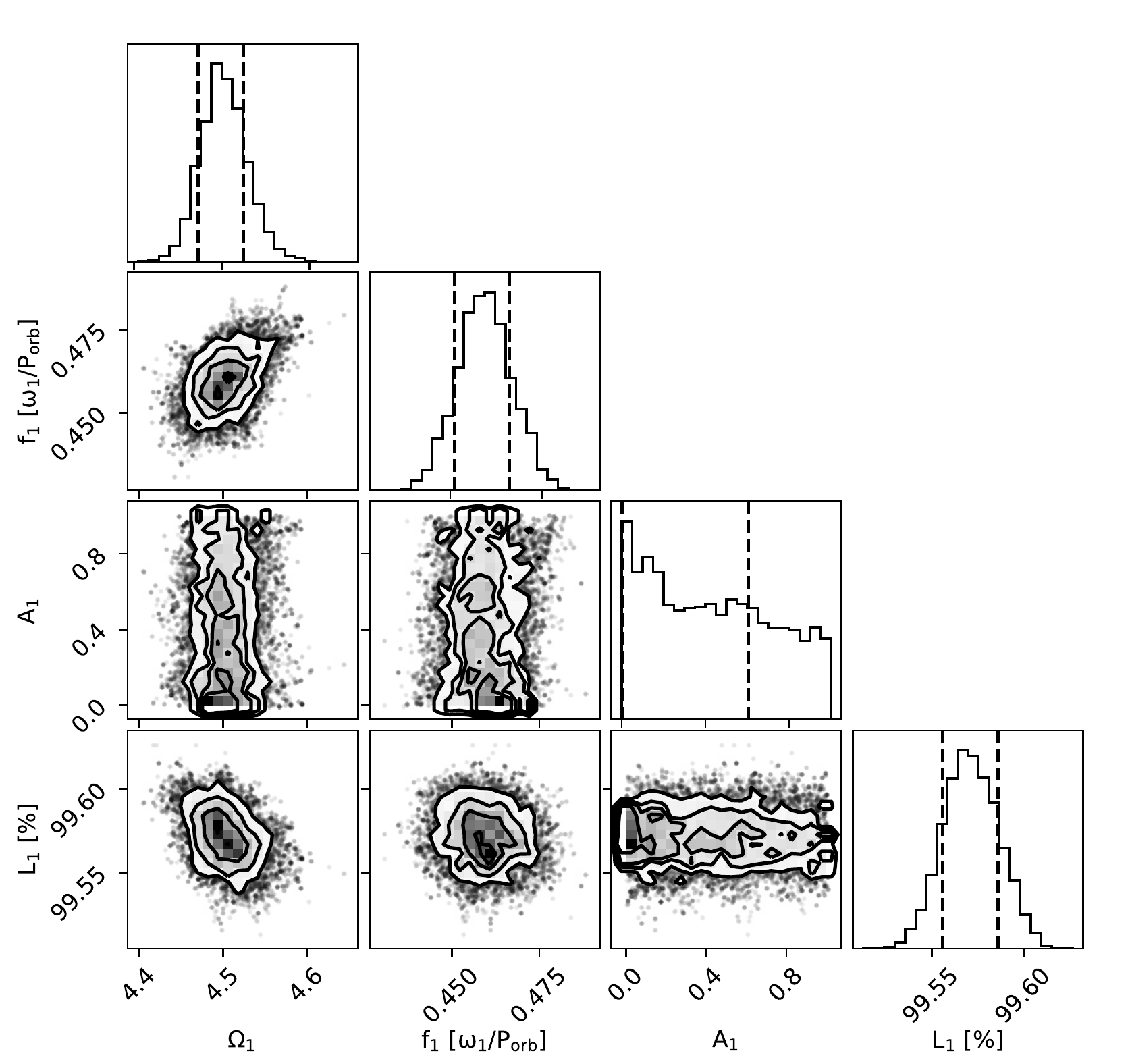}
\caption{Same as Fig.~\ref{fig:pd1}, but for the parameters of the primary star.}
\label{fig:pd3}
\end{figure}

\begin{figure}
\centering
\figps{\hsize}{0}{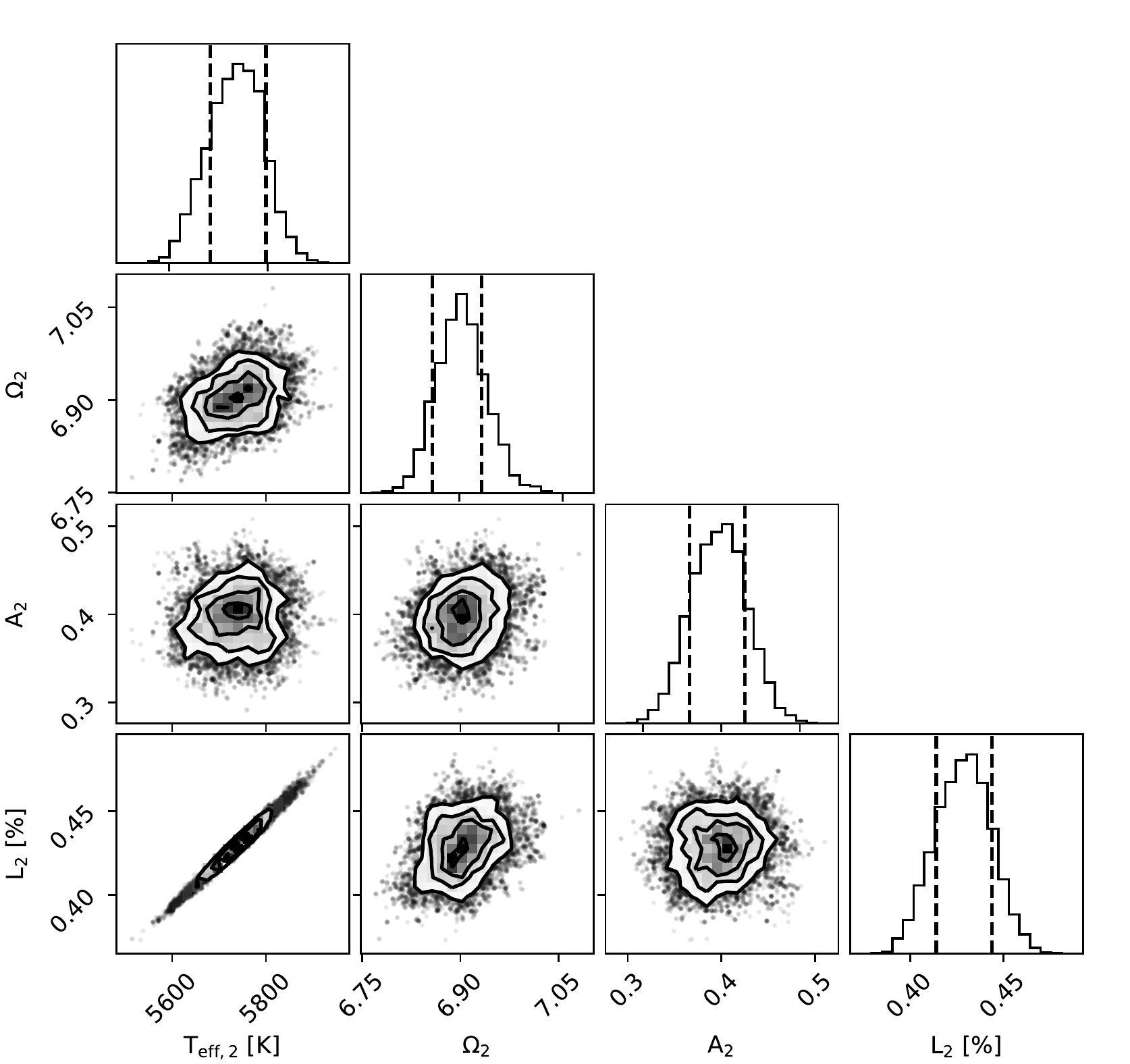}
\caption{Same as Fig.~\ref{fig:pd1}, but for the parameters of the secondary star.}
\label{fig:pd4}
\end{figure}

\section[]{Spectrum synthesis fits and list of lines employed for abundance determination}

\begin{figure*}
\centering
\figps{\hsize}{0}{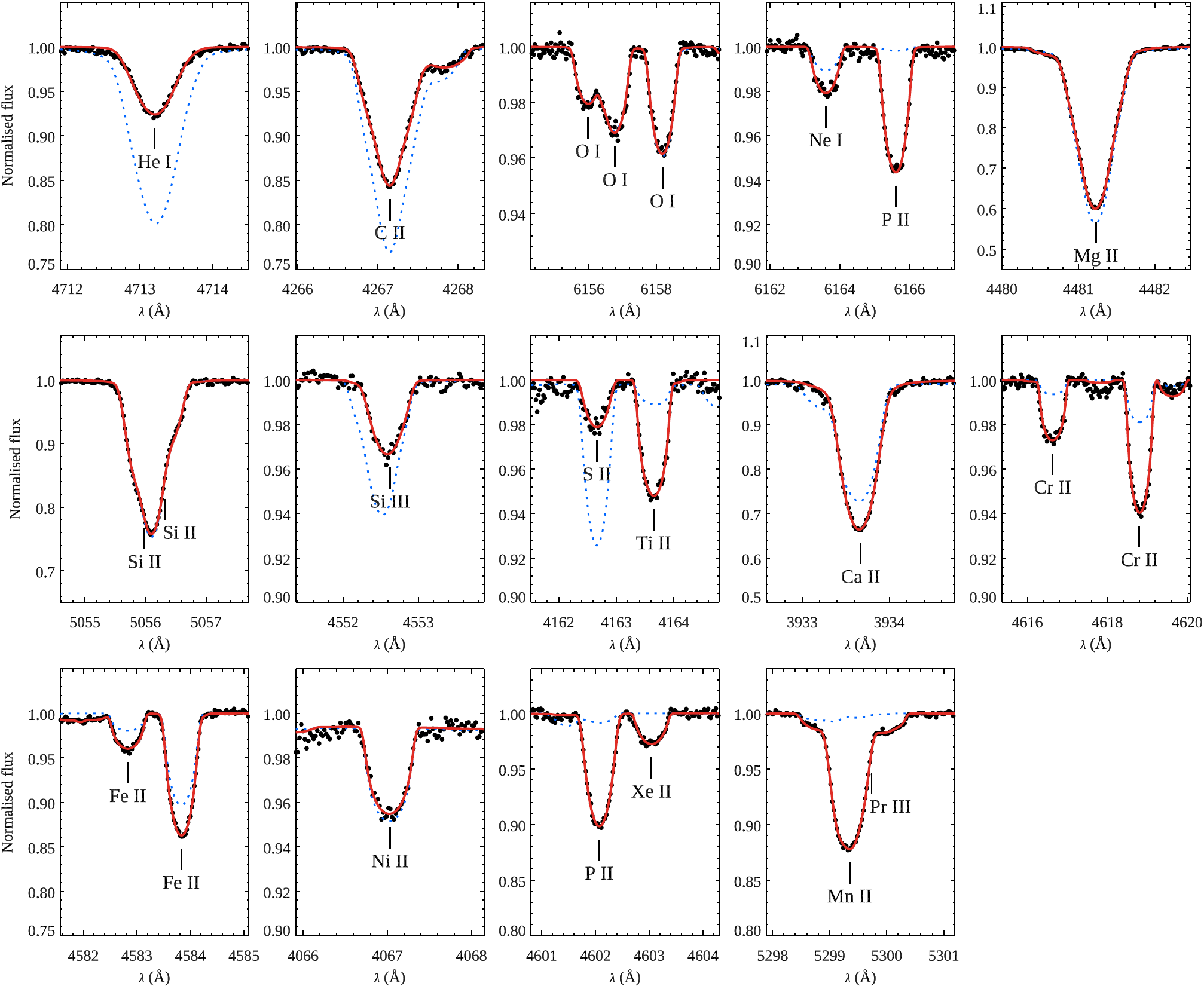}
\caption{Comparison of the average spectrum of \cas\ (symbols) with the best fitting theoretical model spectrum (solid line). The dotted line shows synthetic spectrum calculated with the solar abundances.}
\label{fig:other_fits}
\end{figure*}

\begin{table*}
\caption{Spectral lines employed for determination of abundances in the primary component of \cas. Central wavelengths given with three significant figures correspond to single lines. Wavelengths given with a lower precision indicate lines containing multiple fine, hyperfine and/or isotopic components. \label{tab:lines}}
 \begin{tabular}{ll|ll|ll|ll}
\hline
\hline
Ion & $\lambda$ (\AA) & Ion & $\lambda$ (\AA) & Ion & $\lambda$ (\AA) & Ion & $\lambda$ (\AA) \\
\hline  
\ion{Ti}{ ii} & 3913.461 & \ion{Fe}{ ii} & 4514.516 & \ion{Fe}{ ii} & 4984.465 & \ion{He}{  i} &   5876.6 \\ 
\ion{Ca}{ ii} &   3933.7 & \ion{Fe}{ ii} & 4515.333 & \ion{Fe}{ ii} & 4990.506 & \ion{Si}{ ii} & 5957.559 \\ 
\ion{Fe}{ ii} & 3935.961 & \ion{Fe}{ ii} & 4520.218 & \ion{Fe}{ ii} & 5001.953 & \ion{Fe}{ ii} & 5961.706 \\
\ion{Hg}{ ii} &   3983.9 & \ion{Fe}{ ii} & 4522.628 & \ion{Fe}{ ii} & 5004.188 & \ion{Si}{ ii} & 5978.930 \\
\ion{Mn}{ ii} & 4000.032 & \ion{ P}{ ii} & 4530.823 & \ion{ S}{ ii} & 5014.042 & \ion{ P}{ ii} & 6024.178 \\
\ion{Ni}{ ii} & 4067.031 & \ion{Si}{iii} & 4552.622 & \ion{Fe}{ ii} & 5018.438 & \ion{ P}{ ii} & 6034.039 \\
\ion{He}{  i} &   4120.8 & \ion{Fe}{ ii} & 4555.887 & \ion{Fe}{ ii} & 5035.700 & \ion{ P}{ ii} & 6043.084 \\
\ion{Fe}{ ii} & 4122.659 & \ion{ P}{ ii} & 4558.095 & \ion{Si}{ ii} & 5041.024 & \ion{Ne}{  i} & 6074.338 \\
\ion{Si}{ ii} & 4130.894 & \ion{Cr}{ ii} & 4558.650 & \ion{Si}{ ii} & 5055.984 & \ion{ P}{ ii} & 6087.837 \\                           
\ion{ S}{ ii} & 4162.665 & \ion{Ti}{ ii} & 4563.757 & \ion{Fe}{ ii} & 5061.710 & \ion{Ne}{  i} & 6096.163 \\
\ion{Ti}{ ii} & 4163.644 & \ion{Ti}{ ii} & 4571.971 & \ion{Fe}{ ii} & 5077.896 & \ion{Mn}{ ii} &   6122.4 \\                           
\ion{Mn}{ ii} & 4164.453 & \ion{Fe}{ ii} & 4576.333 & \ion{Mn}{ ii} & 5123.327 & \ion{Mn}{ ii} &   6128.7 \\                           
\ion{Mn}{ ii} &   4205.3 & \ion{Si}{iii} & 4576.849 & \ion{Fe}{ ii} & 5169.028 & \ion{Ne}{  i} & 6143.063 \\                           
\ion{Mn}{ ii} &   4206.3 & \ion{Fe}{ ii} & 4582.830 & \ion{Fe}{ ii} & 5197.568 & \ion{Fe}{ ii} & 6147.734 \\                           
\ion{Mn}{ ii} &   4240.3 & \ion{Fe}{ ii} & 4583.829 & \ion{Fe}{ ii} & 5247.956 & \ion{ O}{  i} &   6155.9 \\                           
\ion{Ga}{ ii} & 4251.154 & \ion{ P}{ ii} & 4589.846 & \ion{Fe}{ ii} & 5260.254 & \ion{ O}{  i} & 6158.187 \\                           
\ion{Mn}{ ii} &   4251.7 & \ion{Cr}{ ii} & 4592.049 & \ion{Fe}{ ii} & 5275.997 & \ion{Ne}{  i} & 6163.594 \\                           
\ion{Mn}{ ii} &   4252.9 & \ion{ P}{ ii} & 4602.069 & \ion{Mn}{ ii} &   5297.0 & \ion{ P}{ ii} & 6165.598 \\                           
\ion{Mn}{ ii} &   4259.2 & \ion{Xe}{ ii} & 4603.040 & \ion{Mn}{ ii} &   5299.3 & \ion{Ne}{  i} & 6266.495 \\                           
\ion{ C}{ ii} &   4267.2 & \ion{Cr}{ ii} & 4616.629 & \ion{Pr}{iii} & 5299.993 & \ion{Ga}{ ii} &   6334.9 \\                           
\ion{Fe}{ ii} & 4273.320 & \ion{Cr}{ ii} & 4618.803 & \ion{Mn}{ ii} &   5302.4 & \ion{Ne}{  i} & 6402.248 \\                           
\ion{Ti}{ ii} & 4290.215 & \ion{Fe}{ ii} & 4620.513 & \ion{ S}{ ii} & 5320.723 & \ion{Ga}{ ii} &   6419.1 \\                           
\ion{Mn}{ ii} &   4292.2 & \ion{Cr}{ ii} & 4634.070 & \ion{ P}{ ii} & 5344.729 & \ion{ P}{ ii} & 6459.945 \\                           
\ion{Fe}{ ii} & 4296.566 & \ion{Fe}{ ii} & 4635.317 & \ion{ S}{ ii} &   5345.9 & \ion{ P}{ ii} & 6503.398 \\                           
\ion{Ti}{ ii} & 4301.922 & \ion{He}{  i} &   4713.1 & \ion{ P}{ ii} & 5409.722 & \ion{Ne}{  i} & 6506.528 \\                           
\ion{Mn}{ ii} &   4326.6 & \ion{Mn}{ ii} &   4727.8 & \ion{Xe}{ ii} & 5419.150 & \ion{ P}{ ii} & 6507.979 \\                           
\ion{Fe}{ ii} & 4351.762 & \ion{Mn}{ ii} &   4730.4 & \ion{Fe}{ ii} & 5427.816 & \ion{ C}{ ii} & 6578.050 \\                           
\ion{Mn}{ ii} &   4356.6 & \ion{Fe}{ ii} & 4731.448 & \ion{Fe}{ ii} & 5429.967 & \ion{ C}{ ii} & 6582.580 \\                           
\ion{Mn}{ ii} &   4363.2 & \ion{Mn}{ ii} &   4755.7 & \ion{ S}{ ii} & 5453.855 & \ion{Ne}{  i} & 6598.953 \\                           
\ion{Mn}{ ii} &   4365.2 & \ion{Mn}{ ii} & 4764.624 & \ion{ S}{ ii} & 5473.614 & \ion{He}{  i} & 6678.154 \\                           
\ion{Ti}{ ii} & 4395.031 & \ion{Mn}{ ii} &   4764.7 & \ion{Fe}{ ii} & 5482.306 & \ion{Ne}{  i} & 6717.043 \\                           
\ion{ P}{ ii} & 4420.712 & \ion{Mn}{ ii} &   4770.3 & \ion{Fe}{ ii} & 5493.831 & \ion{Ne}{  i} & 7032.413 \\                           
\ion{Ti}{ ii} & 4443.801 & \ion{Mn}{ ii} & 4791.762 & \ion{ P}{ ii} & 5499.697 & \ion{He}{  i} &   7065.2 \\                           
\ion{ P}{ ii} & 4452.472 & \ion{Mn}{ ii} &   4806.8 & \ion{Fe}{ ii} & 5506.199 & \ion{Mn}{ ii} &   7330.5 \\                           
\ion{ P}{ ii} & 4463.027 & \ion{Cr}{ ii} & 4824.127 & \ion{Mn}{ ii} &   5559.0 & \ion{Mn}{ ii} &   7353.3 \\                           
\ion{ P}{ ii} & 4475.270 & \ion{Mn}{ ii} & 4830.061 & \ion{Mn}{ ii} &   5570.5 & \ion{Mn}{ ii} &   7367.0 \\                           
\ion{Mn}{ ii} &   4478.6 & \ion{Xe}{ ii} & 4844.330 & \ion{Mn}{ ii} &   5578.1 & \ion{Mn}{ ii} &   7415.8 \\                           
\ion{Mg}{ ii} &   4481.6 & \ion{Fe}{ ii} & 4921.921 & \ion{Fe}{ ii} & 5645.390 & \ion{Mn}{ ii} &   7432.1 \\                           
\ion{Fe}{ ii} & 4489.176 & \ion{ S}{ ii} & 4925.343 & \ion{Mn}{ ii} & 5826.288 & \ion{ P}{ ii} & 7845.613 \\                           
\ion{Fe}{ ii} & 4491.397 & \ion{ P}{ ii} & 4943.497 & \ion{ S}{ ii} &   5840.1 & \ion{Mg}{ ii} & 7877.054 \\ 
\ion{Fe}{ ii} & 4508.280 & \ion{Fe}{ ii} & 4951.581 & \ion{Ne}{  i} & 5852.488 & \ion{Mg}{ ii} & 7896.366 \\
\hline
\end{tabular}
\end{table*}

\label{lastpage}

\end{document}